\documentclass[%
reprint,
superscriptaddress,
amsmath,amssymb,
aps,
prx,
floatfix
]{revtex4-2}

\usepackage{graphicx}% Include figure files
\usepackage{dcolumn}% Align table columns on decimal point
\usepackage{bm}% bold math
\usepackage{xcolor}
\usepackage{multirow}
\usepackage[normalem]{ulem}
\usepackage{braket}
\usepackage{float} 

\usepackage{dcolumn,booktabs}
\newcolumntype{d}[1]{D{.}{.}{#1}}
\newcommand\mc[1]{\multicolumn{1}{c}{#1}} % handy shortcut macro
\usepackage{setspace}
\usepackage{hyperref}
\hypersetup{
    colorlinks=true,
    allcolors=[RGB]{46,48,147}
    }
\usepackage{rotating}
\usepackage[resetlabels,labeled]{multibib}
\newcites{S}{Reference}
\begin{document}

\preprint{APS/123-QED}

\title{Theory and Discovery of Electrides}% 

\author{Chengcheng Xiao}
\email{c.xiao19@imperial.ac.uk}
\affiliation{
Departments of Materials and Physics,
and the Thomas Young Centre for Theory and Simulation of Materials,
Imperial College London, London SW7 2AZ, United Kingdom
}%
\author{Nicholas Bristowe}
\affiliation{Centre for Materials Physics, Durham University, South Road, Durham DH1 3LE, United Kingdom}
\author{Arash A. Mostofi}%
\email{a.mostofi@imperial.ac.uk}
\affiliation{
Departments of Materials and Physics, and the Thomas Young Centre for Theory and Simulation of Materials,
Imperial College London, London SW7 2AZ, United Kingdom
}%

\date{\today}% It is always \today, today,

\begin{abstract}
Electrides are materials with electrons localized at interstitial regions of the crystal lattice and have been identified as promising candidates for a variety of applications, including catalysis, electron emission, and superconductivity. 
We present a theoretical framework for the origin of interstitial electrons in electrides. We demonstrate that this theory can explain electride-like behavior in prototypical electrides, and we use it to develop descriptors for the high-throughput discovery of new inorganic electride candidates from first principles. We also show that the same concepts can explain electride-like behavior in other classes of material, including high-pressure electrides and organic electrides and, more broadly, provide an alternative understanding of F-center defects and solvated electrons.
\end{abstract}

\maketitle

\section{Introduction}
\label{sec:Introduction}

Electrides are materials with electrons localized in the interstitial regions within the crystal structure~\cite{Dye2003}.
These interstitial electrons act as anions and give rise to unique optical and electronic behavior, with potential applications as electron emitters~\cite{Kim2006,Huang1990}, battery anodes~\cite{Hu2015,Hou2016}, catalysts~\cite{Yamamoto2013,Kitano2012,Ye2017}, superconductors~\cite{Zhang2017,Pereira2021,Miyakawa2007} and materials with topological characteristics~\cite{Hirayama2018,Huang2018,Zhang2018}.

Electrides have been observed in many different classes of material. Examples include: the mayenite Ca$_{24}$Al$_{28}$O$_{64}$~\cite{Matsuishi2003}, in which electrons are localized in cages consisting of six neighboring Ca atoms; the insulating high-pressure ($\simeq 300~\mathrm{GPa}$) hP4 phase of elemental sodium~\cite{Ma2009}; the layered materials $\mathrm{Ca_2N}$~\cite{Lee2013} and $\mathrm{Y_2C}$~\cite{Zhang2014,Park2017}, in which electrons are localized in the interlayer planes; Y$_5$Si$_3$, in which electrons are localized along one-dimensional interstitial columns~\cite{Zheng2021}; and organic electrides, such as Na(Tripip222)~\cite{Redko2005} and Cs(18-crown-6)~\cite{Ellaboudy1983}, in which electrons are localized in large ($\simeq 300~\text{\AA}^3$) interstitial regions between the constituent organometallic building blocks.

Despite the fact that numerous materials are recognized as electrides, there is no general-purpose experimental characterization method for unambiguously determining whether or not a material is an electride.
For this reason, electronic structure calculations based on density-functional theory (DFT) have played an important role in identifying electrides via analysis of electron localization~\cite{Liu2020,Dale2018}. 
Descriptors based on electron localization have also been used in high-throughput studies to predict new electrides \cite{Tada2014,Zhou2019,Burton2018}.
Whilst theories of the origin of the electride state have been proposed in the context of high pressure electrides with anomalously small inter-ionic separation~\cite{Miao2014,Miao2015,Dong2017,Modak2019}, there is no consensus about the fundamental theory that underlies why certain materials are electrides and what criteria lead to electron localization at interstitial sites in a crystal at ambient conditions.

In this article, we present a theoretical framework for the origin of interstitial electrons in electrides.
The theory is based on the idea that the localized interstitial electronic states result from multicentered bonding between atomic orbitals of the surrounding atoms.
We demonstrate that this theory can explain electride-like behavior in known prototypical electrides, and we use it to develop descriptors for the high-throughput discovery of new inorganic electride candidates from first principles. 
We apply our automated workflow to $\sim$52,000 entries in the Materials Project database~\cite{Jain2013} to rank these materials according to a figure-of-merit that quantifies the degree to which they are electride-like, which can be used to prioritize more detailed future investigation of specific candidate electrides.
Finally we discuss the wider implications of our theory and show that it can be used to explain organic electrides and provide an alternative understanding of F-center defects and solvated electrons.

\section{Multicentered Bonding Theory}
\label{sec:theory}

We hypothesize that an electronic state localized in an interstitial region of the crystal lattice is a multicentered bonding orbital that is a linear combination of atomic orbitals of the atoms that surround the interstitial region.

\begin{figure}[ht]
    \centering
    \includegraphics[scale=0.8]{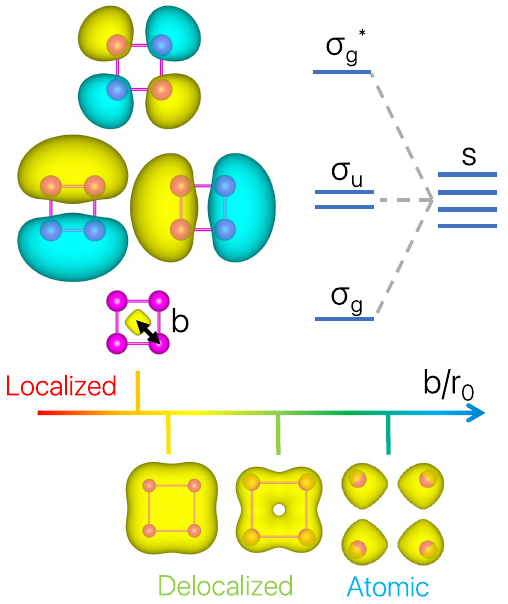}
    \caption{Schematic energy level diagram (upper right panel) and corresponding wavefunction plots (upper left and bottom panels) for a square molecule with an s-orbital on each atom. The form of the lowest energy bonding orbital ($\sigma_\mathrm{g}$) depends on the ratio of the size of the interstitial region (measured by the distance $b$ between the center and the corner atoms) and the size of the orbital (measured by the position $r_0$ of the maximum of its radial distribution function). When this ratio is less than a critical value, $\sigma_\mathrm{g}$ is localized in the interstitial region.
    % \textcolor{orange}{[CCX:just realized that b/r0 is brought up as THE quantity to study, but we never actually use it.]}
    }
    \label{fig:theory-model}
\end{figure}

To illustrate, Fig.~\ref{fig:theory-model} shows the orbital energy level diagram of a model square molecule with one atomic s-orbital on each of the four corner atoms.
Within a non-interacting linear combination of atomic orbitals (LCAO) framework, the lowest-energy bonding molecular orbital is an in-phase combination of atomic orbitals on each site.
Its shape is determined by the ratio of the size of the interstitial region (quantified by the distance $b$ from its center to the nearest surrounding atoms) to the size of the s-orbitals (quantified by the distance $r_{\mathrm{0}}$ from the center of the orbital to the outermost maximum of the radial distribution function of the orbital). When the ratio $b/r_{\mathrm{0}}$ is less than a critical value, $\sigma_g$ is localized in the interstitial region. 
The presence of such an interstitial electronic state, together with its partial occupation, are necessary conditions for the system to be an electride.

Having illustrated the concept of how multicentered bonding can result in localized interstitial electrons in a square cage, we now proceed to generalise the analysis.
We consider interstitial regions defined by $N$ atomic sites of the same elemental species that are arranged in a regular polygon (2D) or polyhedron (3D) around the center of the interstitial site. As above, we denote the distance from the center to each atom by $b$. We consider elements with valence s-electrons, namely those in groups I and II of the periodic table, and determine the condition for the existence of interstitial multicentered bonding \footnote{In principle, our model can be further extended to other atomic orbitals (e.g., p and d) or molecular orbitals such as ``super-atom molecular orbitals'' (SAMOs)~\cite{Feng2008,Feng2011,Johansson2016}, as well as irregular atomic cages.}.
This is done by evaluating the gradient and Hessian of the lowest energy bonding linear combination of the atomic s-orbitals at the interstitial center as a function of the size $b$ of the polyhedron cage, and requiring the multicentered bonding orbital to have a maximum at the interstitial center. Whilst we could do this numerically, using the all-electron atomic radial wavefunctions of each element calculated from density functional theory, we adopt an analytical approach in which the equation for a hydrogenic $n$s-orbital (where $n$ denotes the principal quantum number) is fitted to the tail of the valence all-electron atomic radial wavefunction for each element. Further details are provided in Sec.~\ref{sec:analytical_solutions} of the Supplemental Material (SM)~\cite{[{See Supplemental Material at }][{ for detailed derivation of the critical cage size, DFT details, details of band structure projection schemes, the workflow of the electride descriptor, details of the electron localization function and bonding analysis of more prototypical electride systems.}]supp}. The result is shown in Table~\ref{tab:theory-critical_b}, which lists the critical cage size $b_{\mathrm{c}}$ such that for $b<b_{\mathrm{c}}$ there will be a multicentered bonding orbital localized in the interstitial region.

\begin{table}[h]
    \centering
    \caption{Calculated critical interstitial cage size $b_{\mathrm{c}}$ for 2D and 3D cages with group I and II elements as corner atoms.}
    \begin{ruledtabular}
    \begin{tabular}{cccc}
Group & Element & $b_{\mathrm{c}}^{\mathrm{[3D]}}$ [\AA] & $b_{\mathrm{c}}^{\mathrm{[2D]}}$ [\AA] \\ \hline
\multirow{6}{*}{I} & H	& 1.12  & 0.54  \\
& Li	& 2.71  & 2.25  \\
& Na	& 2.88  & 2.52  \\
& K	    & 3.47  & 3.15  \\
& Rb	& 3.63  & 3.35  \\
& Cs	& 3.94  & 3.70  \\
\hline
\multirow{5}{*}{II} & Be	& 1.80  & 1.48  \\
& Mg	& 2.17  & 1.90  \\
& Ca	& 2.74  & 2.50  \\
& Sr	& 2.95  & 2.73  \\
& Ba	& 3.27  & 3.07  \\
    \end{tabular}
    \end{ruledtabular}
    \label{tab:theory-critical_b}
\end{table}

The value of $b_{\mathrm{c}}$ depends on the elemental species of the atoms forming the cage and on whether the cage is 3D or 2D, but not on the number of atomic sites $N$, which is a special feature of the analytical form of hydrogenic s-orbitals and our consideration of regular cages. Also, it can be seen that $b_{\mathrm{c}}$ increases as the principal quantum number of the valence s-electrons increases, and decreases when going from group I to group II. These qualitative trends can be understood in terms of the change in size of the valence s-orbitals. The values of $b_{\mathrm{c}}$ are typical of the size of interstitial regions in many real materials ($\sim 2$ to $3$~\AA), even at ambient conditions. Despite the simplicity of the LCAO model, as we shall see when we analyze specific electrides in Sec.~\ref{sec:validation}, $b_\mathrm{c}$ can provide a valuable heuristic for indicating whether a material might host a multicentred bonding orbital in an interstitial region.

\section{Validation on known electrides}
\label{sec:validation}
 
We now demonstrate that the localized interstitial electrons in the known electrides $\mathrm{Ca_2N}$ and hP4-Na originate from multicentered bonding combinations of atomic s-orbitals, as described by the model in Sec.~\ref{sec:theory}.

\begin{figure*}[ht]
    \centering
    \includegraphics[scale=1]{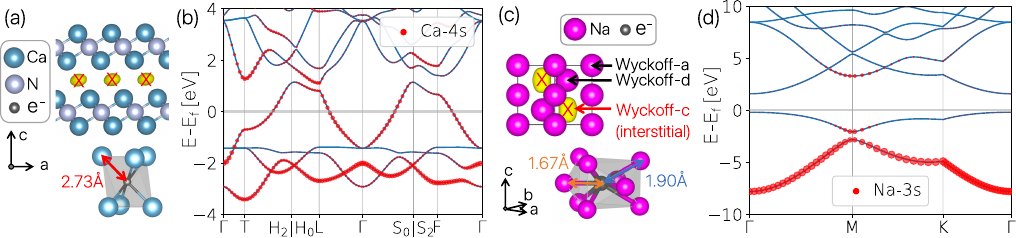}
    \caption{(a) Upper panel: Electronic charge density associated with the band that crosses the Fermi level in $\mathrm{Ca_2N}$. It is localized between two layers at the center of an octahedron formed by Ca atoms, marked by the red crosses. Lower panel: zoom-in of the Ca octahedron. (b) Bandstructure of $\mathrm{Ca_2N}$ projected onto Ca 4s-orbitals. The size of each red circle indicates the magnitude of the projection. (c) Upper panel: Electronic charge density associated with the band located between -8~eV and -2eV in hp4-Na. The centre of the interstitial cage is located at Wyckoff site $c$. Lower panel: zoom-in of the interstitial cage formed by the nearest and second-nearest Na atoms of the interstitial site. (d) Bandstructure of hP4-Na projected onto Na 4s-orbitals. The size of each red circle indicates the magnitude of the projection.}
    \label{fig:theory-validation}
    % NOTE change charge density are the red band that crosses the Fermi energy, including the unoccupied states.
    \end{figure*}

$\mathrm{Ca_2N}$ is a layered material whose atomic structure, shown in the top panel of Fig.~\ref{fig:theory-validation}(a), is similar to the T-phase of transition metal dichalcogenides. The stacking configuration is such that the Ca atoms in one layer are directly above/below the N atoms in an adjacent layer~\cite{Lee2013}. 

We fully relax the atomic structure and lattice parameters within density-functional theory with additional dispersion correction in the form of Tkatchenko-Scheffler \cite{Tkatchenko2009} with iterative Hirshfeld partitioning \cite{Bultinck2007} to account for the van der Waals interactions between layers. Further details of the calculation method and relaxed lattice parameters and atomic coordinates are given in Sec.~\ref{sec:Methods} of the SM~\cite{[{See Supplemental Material at }][{}]supp}.

The octahedral interstitial sites with Ca atoms at the corners and located between $\mathrm{Ca_2N}$ layers (Fig.~\ref{fig:theory-validation}(a), bottom panel) have a cage size $b=2.73$~\AA, which is comparable to the critical value for Ca in Table~\ref{tab:theory-critical_b} ($b_{\mathrm{c}}^{[3\mathrm{D}]}=2.74$~\AA) and indicates that this material is a potential candidate for hosting an interstitial localized electron.

Fig.~\ref{fig:theory-validation}(b) shows the projected band structure using PAW pseudo partial waves (see Sec.~\ref{sec:proj_PAW_partial_wave} of the SM~\cite{[{See Supplemental Material at }][{}]supp} for further details). The state that crosses the Fermi energy is almost entirely composed of Ca 4s-orbitals.
The charge density associated with this partially occupied band is shown in the top panel of Fig.~\ref{fig:theory-validation}(a). 
It can be clearly seen that the partial charge density is localized (the maxima are marked with a red ``X'') in the octahedral interstitial regions between the $\mathrm{Ca_2N}$ layers, which is consistent with this material's identification as an electride~\cite{Lee2013}. 

Fig.~\ref{fig:theory-validation}(c) shows the hP4 phase of sodium (hP4-Na). This is a double hexagonal closed-packed structure (space group P6$_3$/mmc) that is stable at a hydrostatic pressure of 320~GPa~\cite{Ma2009}.
There are four Na atoms in the primitive cell: two on the edge (Wyckoff site $a$) and two inside the primitive cell (Wyckoff site $d$). There are also two interstitial sites (Wyckoff site $c$), each with three first nearest neighbor Na atoms forming a triangle with $b=1.67$~\AA~and six second nearest neighbors forming a triangular prism with $b=1.9$~\AA\ (Fig.~\ref{fig:theory-validation}(c), bottom left). Details of the relaxed structure and calculation method are provided in Sec.~\ref{sec:Methods} of the SM~\cite{[{See Supplemental Material at }][{}]supp}.
These interstitial regions are much smaller than the critical cage size for Na in Table~\ref{tab:theory-critical_b} ($b_{\mathrm{c}}^{[3\mathrm{D}]}=2.167$~\AA), indicating the potential for an electride-like interstitial state arising from multicentered bonding in these cages.

Fig.~\ref{fig:theory-validation}(d) shows the band structure projected onto the 4s-orbitals of Na. The fully occupied bonding orbital between $-8$~eV and $-2$~eV is almost entirely composed of Na 4s-orbitals. 
The charge density associated with this band, shown in Fig.~\ref{fig:theory-validation}(c), demonstrates it to be localized within the interstitial regions discussed above, confirming that the interstitial electronic state in hP4-Na may be understood in terms of the multicentered bonding theory presented above.

It is worth noting that the fully occupied top valence band also corresponds to an interstitial localized electronic state that is composed of Na atomic 3p-orbitals (see Sec.~\ref{sec:bonding_analysis} of the SM).
% \textcolor{blue}{[AAM: change SI to SM, and refer to the relevant section in the same way as we have done in earlier examples, eg, ``see Sec. XX of the SM''.]}
This suggests that it is also important to consider multicentered bonding that arises from atomic orbitals with higher angular momentum (i.e., p, d and f orbitals). Whilst in principle the analytical model may be extended to such cases, a more general framework based on first-principles descriptors is desirable and developed in Sec.~\ref{sec:descriptor} below. 

Finally, we have performed detailed analysis on several other known electride systems that further confirms our hypothesis that electrides originate from multicentered bonding and which can be found in Sec.~\ref{sec:bonding_analysis} of the SM~\cite{[{See Supplemental Material at }][{}]supp}.

\section{Descriptor for electrides}
\label{sec:descriptor} 

Whilst the analytical LCAO model presented in Sec.~\ref{sec:theory} provides intuitive insight into the origin of multicentered bonding in materials and the associated existence of interstitial localized electrons, its main shortcomings are that it doesn't permit prediction of whether or to what extent these orbitals are occupied, which is another necessary condition for a material to be an electride, and that it doesn't take into account the real interacting Hamiltonian of the system. This encourages us to take a more general approach based on first-principles calculations that can identify \emph{occupied} multicentered bonding orbitals in materials. 

The electron localization function (ELF) is a scalar field that can be calculated from the occupied eigenstates obtained from an electronic structure calculation. It quantifies the degree of Pauli exclusion of electrons at each point in space, with a value ranging from zero to one, whereby larger values indicate greater localization of electrons~\cite{Burdett1998,Becke1990} (see Sec.~\ref{sec:ELF} of the SM~\cite{[{See Supplemental Material at }][{}]supp} for a review of the electron localisation function and how its value should be interpreted).
As such, the ELF has been widely used to identify the location of occupied covalent bonds~\cite{Grin2014}.
But the ELF also provides a natural way to identify the existence and location of occupied multicentered bonding orbitals~\cite{Grin2014} and, thereby, to identify potential electride materials.
Indeed, the ELF has been discussed and used recently as an appropriate tool for helping to identify electrides~\cite{Dale2018, Zhao2016, Wan2018,Zhou2019,Zhang2017prx, Zhang2014, Lee2013, Dong2017nc}, although the underlying connection to multicentered bonding was not made.

\begin{figure}[ht]
    \centering
    \includegraphics[scale=1]{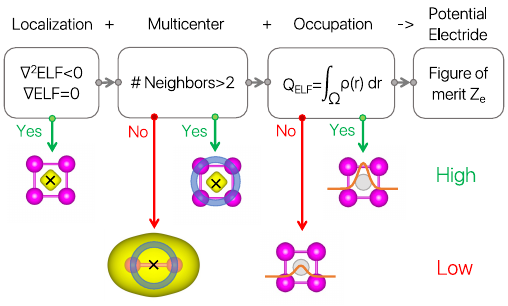}
    \caption{Schematic workflow for the electride figure of merit $Z_\mathrm{e}$. First, we determine the positions of the local maxima of the ELF by computing its gradient and Hessian. Second, we determine whether any of these local maxima might be associated with multicenter bonding by computing the distance to shells of nearest neighbor atoms of the same elemental species, and selecting those with more than two atoms in the same shell. Third, we determine the total electronic charge $Q_\mathrm{ELF}$ associated with an ELF local maximum by Bader partitioning the ELF and integrating the electronic charge density of the system over the Bader-like basin $\Omega$ associated with it. Finally, the value of the ELF at the local maximum $\mathrm{ELF}(\vec{r}_{\mathrm{c}})$ and its associated $Q_\mathrm{ELF}$ are used to compute the electride figure of merit $Z\mathrm{_e}$ (Eq.~\ref{eq:Ze}).}
    \label{fig:theory-descriptor}
\end{figure}

Here, in the first step of our approach, we identify multicentered bonds by analyzing the topology of the ELF. We do this by searching for local maxima of the ELF that are situated within interstitial regions of the crystal structure. In practice, we make this determination by considering the distance from the position of the ELF local maximum to the shell of nearest neighbor atoms of the same elemental species.
If there are only two atoms in the first nearest neighbor shell, then we identify the ELF maximum as corresponding to a covalent bond; if there are more than two atoms in the first nearest neighbor shell, however, then we identify the ELF maximum as corresponding to the site of a multicentered bonding orbital in an interstitial region. These steps are shown schematically as the first two boxes in Fig.~\ref{fig:theory-descriptor}. 

The second step of our approach is to quantitatively determine the occupation of the multicentered bonds identified above. We achieve this via topological analysis based on Bader's Quantum Atoms in Molecules (QTAIM) method~\cite{Bader1991}, but applied to the ELF rather than the electronic charge density. This partitions the crystal into Bader-like basins according to the zero-flux planes of the ELF. Once these ELF basins have been determined, those that are associated with the ELF local maxima that have been identified in the first step above can be used as integration volumes for the electronic charge density. In this way, an estimate of the electronic charge associated with a basin around a given ELF maximum, and hence a given multicentered bonding orbital, is calculated.
This second step is represented by the third box in Fig.~\ref{fig:theory-descriptor}, which shows the complete workflow of our approach to identifying electrides. A more detailed description of the workflow is provided in Sec.~\ref{sec:Search_protocol} of the SM~\cite{[{See Supplemental Material at }][{}]supp}.

The value of the ELF at a multicentered orbital centre $\vec{r}_{\mathrm{c}}$ and the integrated charge $Q_{\mathrm{ELF}}$ associated with its Bader-like ELF basin are heuristically combined into an electride figure of merit $Z_\mathrm{e}$, given in the non-spin-polarized case by
\begin{equation}
	Z_\mathrm{e} = \frac{1}{2}\left[ \mathrm{ELF}(\vec{r}_{\mathrm{c}}) + \frac{Q_\mathrm{ELF}}{2} \right].
	\label{eq:Ze}
\end{equation}
The value of $Z_\mathrm{e}$ ranges from zero to one, with a larger value indicating the system is more likely to be a potential electride by having one or both a higher value of ELF at the multicentered bonding orbital centre and a higher occupancy of this orbital. $Z_\mathrm{e}$ can therefore be used as a descriptor to rank potential candidate electrides.

\section{Screening of electrides}
\label{sec:results} 

We use the descriptors and workflow presented in Sec.~\ref{sec:descriptor} to perform automated high-throughput first-principles screening for potential electrides within the 51,913 crystal structures in the Materials Project database. Technical details of the density-functional theory calculations are provided in Sec.~\ref{sec:Methods} of the SM~\cite{[{See Supplemental Material at }][{}]supp}.

We find around 10,000 potential candidate materials that possess one or more local interstitial ELF maxima that are indicative of multicentered bonding. 

\begin{figure}[ht]
    \centering
    \includegraphics[scale=0.9]{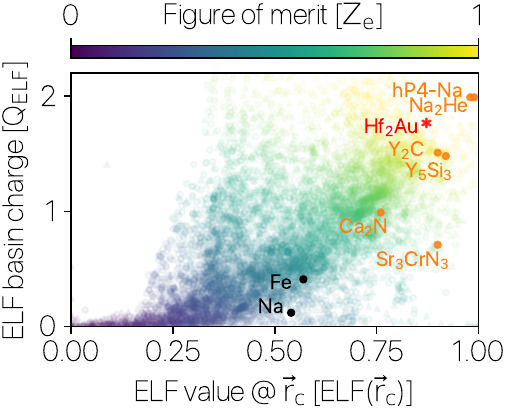}
    \caption{Scatter plot of the ELF basin charge $Q_\mathrm{ELF}$ and ELF values at the cage center ELF($\vec{r}_{\mathrm{c}}$) for the $\sim$10,000 systems identified by our screening workflow (Fig.~\ref{fig:theory-descriptor}). 
    % \textcolor{red}{Circles represent non-magnetic materials and triangles represent materials with potential magnetic ground states} \textcolor{blue}{[AAM: Maybe remove this as you can't really see circles or triangles in the plot?]}. 
    The figure of merit $Z_\mathrm{e}$ (Eq.~\ref{eq:Ze}) for each material is indicated by the color of the data point. Some known electrides, non-electrides, and a new potential electride Hf$_2$Au with high $Z_\mathrm{e}$ are highlighted (see also Table~\ref{tab:results-FoM}).}
    \label{fig:results-FoM}
\end{figure}

Fig.~\ref{fig:results-FoM} shows a scatter plot of the value of the ELF at the local ELF maximum and the ELF basin charge $Q_{\mathrm{ELF}}$ for these systems, where the color of each data point reflects the value of the electride figure of merit $Z_{\mathrm{e}}$ (Eq.~\ref{eq:Ze}).
There is a clear correlation between the ELF value and basin charge. This is not entirely surprising as the ELF is a measure of the localization of the occupied states. 
There is no clear demarcation, however, between electrides and non-electrides. This is consistent with the theory presented in Sec.~\ref{sec:theory} in the sense that the form of a multicentered bonding orbital evolves continuously as a function of the structure of the interstitial cage (Fig.~\ref{fig:theory-model}), and the occupancy of such an orbital depends on the chemical details of the system.
For example, many of the points plotted in Fig.~\ref{fig:results-FoM} with a relatively low figure of merit ($Z_{\mathrm{e}}\lesssim 0.6$) correspond to intermetallic systems. Several of these (e.g., Yb$_5$Sb$_3$ and Sr$_3$Li$_2$) have been the subject of debate as to whether they should be classified as electrides~\cite{Nesper1991,Liu2020}. 
If we take the figure of merit as an indicator of a material's potential for being an electride, this suggests that there are many other candidate materials that have stronger electride characteristics.
Indeed, almost all previously identified electrides that are present in the Materials Project database, such as $\mathrm{Y_5Si_3}$, $\mathrm{Y_2C}$, $\mathrm{Sr_3CrN_3}$ and $\mathrm{Ca_2N}$, as well as the high-pressure electrides hP4-Na and $\mathrm{Na_2He}$, typically have high $Z_{\mathrm{e}}$ and are listed in the middle panel of Table~\ref{tab:results-FoM}. Conversely, conventional non-electride materials either don't have an interstitial local maximum in the ELF, or have very low figure of merit; a selection of such systems is listed in the bottom panel of Table~\ref{tab:results-FoM}. This gives us confidence in our screening protocol.

A significant number of additional systems have a high electride figure of merit, and are good candidates for further investigation as potential new electrides. Restricting ourselves to materials that have been experimentally synthesized, we list the top 10 in the upper panel of Table~\ref{tab:results-FoM}.

\begin{table}[!ht]
    \centering
    \caption{Calculated ELF local maxima value ($\mathrm{ELF}(\vec{r}_{\mathrm{c}})$), ELF basin charge ($Q_\mathrm{ELF}$) and electride figure of merit ($Z_\mathrm{e}$) for 10 experimentally synthesized systems from our screening process, six known electrides and six non-electride materials, ranked by their $Z_\mathrm{e}$.}
    \begin{ruledtabular}
    \begin{tabular}{ccccc}
Type & Formula      & $\mathrm{ELF}(\vec{r}_{\mathrm{c}})$ & $Q_\mathrm{ELF}$ & $Z_\mathrm{e}$   \\ \hline
 \multirow{10}{*}{\begin{sideways}Screening results\end{sideways}} &Eu$_5$As$_3$~\cite{Wang1978} & 0.88   & 1.99             & 0.94     \\ % mp-1106024         
 &Eu$_5$Bi$_3$~\cite{Corbett2006} & 0.86   & 1.92             & 0.91     \\ % mp-1188517        
 &Eu$_5$Sb$_3$~\cite{Corbett2006} & 0.87   & 1.87             & 0.90     \\ % mp-1105949 
 &Ca$_2$Sb~\cite{Eisenmann1973}     & 0.98   & 1.60             & 0.89     \\ % mp-9925   
 &Hf$_2$Au~\cite{Schubert1962}     & 0.87   & 1.76             & 0.87     \\ % mp-30383   
 &TiAu~\cite{Schubert1962}         & 0.79   & 1.92             & 0.87     \\ % mp-998972  
 &LaScSb~\cite{Nuss2014}       & 0.84   & 1.76             & 0.86     \\ % mp-1076970 
 &Y$_3$Pt~\cite{LeRoy1979}      & 0.86   & 1.70             & 0.86     \\ % mp-7343    
 &Be~\cite{Martin1959}           & 0.76   & 1.90             & 0.86     \\ % mp-87      
 &NdScSb~\cite{Nuss2014}       & 0.88   & 1.63             & 0.85     \\ % mp-1077443 
\hline
\multirow{6}{*}{\begin{sideways}\parbox{2cm}{Known\\ electrides}\end{sideways}} &hP4-Na~\cite{Ma2009}        & 0.99         & 1.99            &0.99   \\ 
&Na$_2$He~\cite{Dong2017nc}      & 0.98         & 1.99            &0.99   \\
&Y$_5$Si$_3$~\cite{Lu2016a}   & 0.92         & 1.48            &0.83   \\ 
&Y$_2$C~\cite{Zhang2014}       & 0.90         & 1.51            &0.83   \\
&Sr$_3$CrN$_3$~\cite{Chanhom2019} & 0.90         & 0.71            &0.63   \\
&Ca$_2$N~\cite{Lee2013}       & 0.76         & 0.99            &0.63   \\ 
\hline
\multirow{6}{*}{\begin{sideways}\parbox{2cm}{Non-electrides}\end{sideways}} &Fe              & 0.54         & 0.12            &0.30   \\ 
&Na              & 0.57         & 0.41            &0.39   \\
&Si              & N.A.          & N.A.             &N.A.   \\ 
&C(graphite)     & N.A.          & N.A.             &N.A.   \\ 
&NaCl            & N.A.          & N.A.             &N.A.   \\ 
&MgO             & N.A.          & N.A.             &N.A.   \\ 
    \end{tabular}
    \end{ruledtabular}
    \label{tab:results-FoM}
\end{table}

\begin{figure}[ht]
    \centering
    \includegraphics[scale=1]{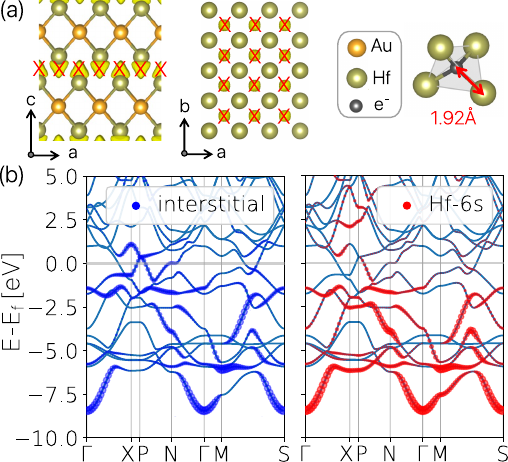}
    \caption{(a) Left panel: ELF of $\mathrm{Hf_2Au}$, looking down the b and c axes respectively. The tetrahedral interstitial sites, formed by Hf atoms in adjacent layers, are marked by red crosses. Right panel: zoom-in of an interstitial cage, which has a size $b=1.92$~\AA. (b) Bandstructure of $\mathrm{Hf_2Au}$ projected onto hydrogenic 1s-orbitals located at the interstitial site centres (left panel) and onto Hf 6s-orbitals (right panel). The size of each red circle indicates the magnitude of the projection. }
    % \textcolor{blue}{[AAM: X and P points in the bandstructure plots are very close. The right-hand plot does not have a vertical line for the P point, whereas the left hand plot does.]}}
    \label{fig:results-Hf2Au}
\end{figure}

As an example, we focus on Hf$_2$Au ($\mathrm{ELF}(\vec{r}_{\mathrm{c}})=0.87$, $Q_{\mathrm{ELF}}=1.76$ and $Z_{\mathrm{e}}=0.87$), a layered material with space group I4/mmm. 
A single layer of Hf$_2$Au consists of a square lattice of Au atoms sandwiched between two square lattices of Hf atoms that are shifted along the [110] direction with respect to the Au layer.
Fig.~\ref{fig:results-Hf2Au}(a) [left panels] shows the atomic structure as well as an ELF isosurface. The ELF is localized in tetrahedral interstitial sites that are formed by Hf atoms from adjacent Hf$_2$Au layers and which have a cage size of $1.92$~\AA\ [see right panel of Fig.~\ref{fig:results-Hf2Au}(a)]. This is less than the critical cage size $b_{\mathrm{c}}^{[3\mathrm{D}]}=2.37$~\AA~that we calculate from our LCAO model for the 6s-orbitals of Hf and, therefore, indicative of a potential interstitial multicentred bond.

Fig.~\ref{fig:results-Hf2Au}(b) shows the bandstructure projected onto a set of pseudo hydrogenic 1s-orbitals positioned at the ELF maxima. We see that the interstitial states can be attributed to two energy regions, one near $-7.5$~eV and the other near the Fermi level. To confirm that these states originate from the atomic 6s-orbitals of the Hf atoms forming the interstitial cage, we also plot in Fig.~\ref{fig:results-Hf2Au}(c) the bandstructure projected onto Hf's 6s PAW pseudo partial waves. We can see that the same bands are highlighted, confirming our expectation.

\section{Organic and other electrides}
\label{subsec:other_electrides} 

Thus far, we have considered localized multicentered bonding orbitals that arise from linear combination of atomic orbitals that surround interstitial regions in a crystal lattice. 
We can, however, consider a similar mechanism based on linear combination of so-called ``super-atom molecular orbitals'' (SAMOs)~\cite{Feng2008,Feng2011,Johansson2016}. 
Metal-organic molecular crystals, such as Cs(15-Crown-5)$_2$ and Na-Tripip222 exhibit localized electrons within large interstitial regions in the crystal structure. Since SAMOs are much larger than atomic orbitals, the critical cage size $b_{\mathrm{c}}$ for the formation of a localized interstitial multicentered bonding orbital is also much larger and corresponds to the typical interstitial cage size found in these systems. 
Further details and analysis of Cs(15-Crown-5)$_2$ and Na-Tripip222 including the ELF, the calculated critical cage size and the RDF of the SAMOs are provided in  Sec.~\ref{sec:organic_electrides} of the SM~\cite{[{See Supplemental Material at }][{}]supp}.

Furthermore, non-periodic systems that possess suitable cages are also expected to exhibit similar phenomena.
For example, F-center defects such as oxygen vacancies in MgO have interstitial localized electrons at the defect center~\cite{Kulichenko2020}, which is similar to an interstitial site and is surrounded by a cage of Mg atoms with $b \simeq 2.12$~\AA. This comparable to the predicted critical cage size of Mg $b_{c}^{\text{[3D]}} = 2.17$~\AA~(see Table~\ref{tab:theory-critical_b}) and, therefore, indicates that these localized electronic states potentially originate from multicentered bonding of the surrounding Mg atomic orbitals. 
This is verified by projection analysis of the bandstructure of the defect supercell; details are given in Sec.~\ref{subsec:other_electrides;F-center} of the SM~\cite{[{See Supplemental Material at }][{}]supp}.
%\textcolor{red}{AAM: This is verified by projection analysis of the bandstructure of the defect supercell; details are given in Sec.~\ref{subsec:other_electrides;F-center} of the SM~\cite{[{See Supplemental Material at }][{}]supp}.}

%
As a final example, first-principles calculations have demonstrated that hydrogen atoms in liquid water form dynamic, irregular polyhedral cages as small as 1~\AA\ that can host localized electrons~\cite{Lan2021}. 
Our predicted critical cage size for hydrogen is comparable, $b_c^\text{[3D]} = 1.06$~\AA\ (see Table~\ref{tab:theory-critical_b}), and, therefore, we speculate that the electronic states associated with these ``solvated electrons'' may also be understood as interstitial multicentered bonds.

% \textcolor{red}{Our findings suggest that interstitial localized electrons are prevalent among material and molecular systems and that multicentered bonding may provide a unified explanation.} \textcolor{blue}{[AAM: I don't think we need this sentence here as it is essentially repeated in the conclusions just a few lines below.]}

\section{Conclusion}
\label{sec:Conclusion}

In conclusion, we have presented a theoretical framework for the origin of interstitial electrons in electrides. We demonstrate that such electronic states can be understood in terms of multicentered bonding between atomic orbitals of the surrounding atoms.
We have validated the theory on known electride systems and have used it to develop an automated high-throughput first-principles workflow based on the electron localisation function (ELF) for identifying potential new electrides from the Materials Project database. 
We rank materials according to a figure-of-merit that can be used to prioritize further detailed investigation of candidate electrides.
Moreover, we have shown wider implications of multicentered bonding to other systems, including organic electrides, F-center defects and solvated electrons.

\begin{acknowledgments}
We are grateful to the UK Materials and Molecular Modelling Hub for computational resources, which is partially funded by EPSRC (EP/P020194/1 and EP/T022213/1).
We also acknowledge computational resources and support provided by the Imperial College Research Computing Service (\href{http://doi.org/10.14469/hpc/2232}{http://doi.org/10.14469/hpc/2232}). The authors further acknowledge the Chinese Scholarship Council for funding via a PhD scholarship.
\end{acknowledgments}

% \appendix
% \nocite{*}

\bibliography{bibliography/references}

%%%%%%%%%% Merge with supplemental materials %%%%%%%%%%
 \widetext
 \begin{center}
 \newpage 
% \resetlinenumber
\begin{spacing}{1.6}
 \textbf{\large --Supplemental Material--} \\
 \textbf{\large Theory and Discovery of Electrides} \\
\end{spacing}

\begin{spacing}{1.3}
\text{Chengcheng Xiao,$^1$ Nicholas Bristowe,$^2$ and Arash A. Mostofi$^1$} \\
\end{spacing}

\textit{$^1$Departments of Materials and Physics, and the Thomas Young Centre for Theory and Simulation of Materials,} \\ 
\textit{Imperial College London, London SW7 2AZ, United Kingdom} \\
\textit{$^2$Centre for Materials Physics, Durham University,} \\
\textit{South Road, Durham DH1 3LE, United Kingdom}

\end{center}

%%%%%%%%%% Merge with supplemental materials %%%%%%%%%%
%%%%%%%%%% Prefix a "S" to all equations, figures, tables and reset the counter %%%%%%%%%%
 \setcounter{equation}{0}
 \setcounter{figure}{0}
 \setcounter{table}{0}
 \setcounter{page}{1}
 \setcounter{section}{0}
 \makeatletter
 \renewcommand{\theequation}{S\arabic{equation}}
 \renewcommand{\thefigure}{S\arabic{figure}}
 \renewcommand{\thetable}{S\arabic{table}}
 \renewcommand{\bibnumfmt}[1]{[S#1]}
 \renewcommand{\citenumfont}[1]{S#1}
%%%%%%%%%% Prefix a "S" to all equations, figures, tables and reset the counter %%%%%%%%%%

 \section{Model to determine critical cage size}
\label{sec:analytical_solutions}
\begin{figure}[H]
     \centering
     \includegraphics[scale=0.8]{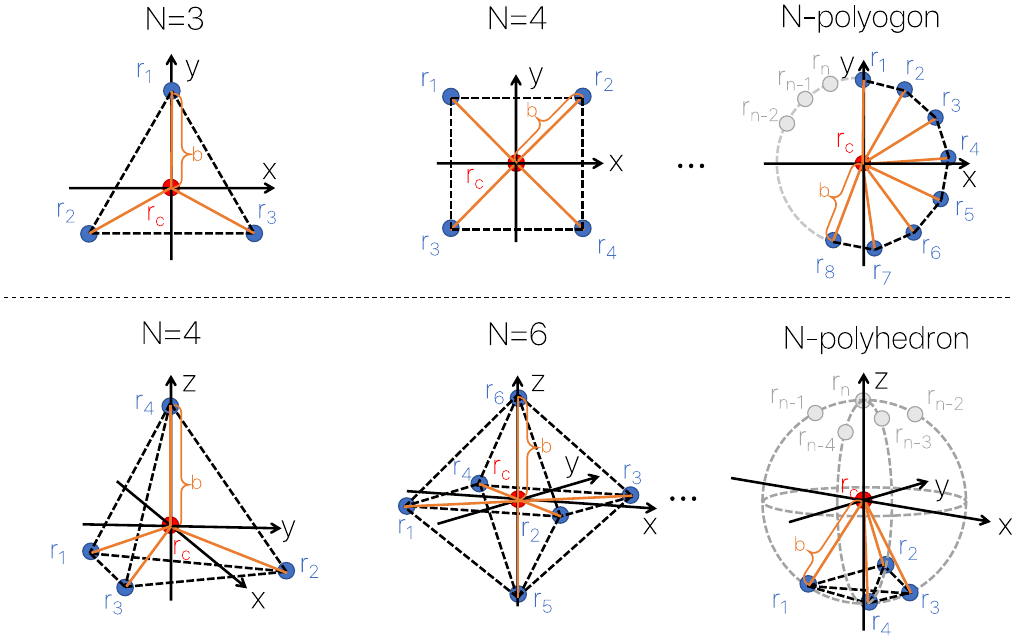}
     \caption{Illustration of the 2D regular polygon models (upper panel) and 3D regular polyhedron models (lower panel). Without loss of generality, the origin is defined as the center $\vec{r}_{\mathrm{c}}$ of each polygon or polyhedron. The distance between the center of the polygon or polyhedron and the position $\vec{r}_n$ of each corner atom is denoted $b$.}
     % Lower panel (top): the radial distribution function $r\psi(r)$ of 1s, 2s and 3s orbitals. Lower panel (bottom): the corresponding eigenvalues of the Hessian at $\vec{r}_{\mathrm{c}}$ for polygons with different number of corners $N$ (bottom). The red shaded regions represent the values of $b$ where the system has a maxima at $\vec{r}_{\mathrm{c}}$. The orange vertical lines highlight the position of the outermost maximum of the radial distribution function for each $n$s-orbital, which we take as our measure of the size $r_0$ of the corresponding orbital. 
     % \textcolor{blue}{[AAM: I would be tempted to do a basis transformation on Figs S1 and S2: make a new figure S1 with just the polygons and polyhedra (ie, showing the geometry of the model in both 2D and 3D), and a new figure S2 that shows the results of the model for both 2D and 3D. You can then refer to S1 in the first paragraph when introducing the model, and S2 later when you talk about the solution/results.]}}
     \label{fig:2D-model}
\end{figure}

%As the smallest molecule, the H$_2$ doesn't have a maxima in its wavefunction at the center of two atoms, regardless of the bond length, due to the exponential decaying feature of Hydrogen's 1s orbital.
%This is a well known property of the hydrogen covalent bond and people have proposed several ways to improve the ELF so that it can still be used to predict the existence of this covalent bond[REF].
%To ensure that this features does not affect the electride-like multicentered bonds, and 
As discussed in the main manuscript, within the non-interacting LCAO approximation, whether or not there is an interstitial localized electronic state depends on the size of the interstitial cage $b$ and size $r_0$ of the atomic s-orbitals.
To understand this dependence in more detail, we study the case in which the interstitial cages are regular polygons (in 2D) and polyhedrons (in 3D) with atomic sites at the corners (Fig.~\ref{fig:2D-model}). We consider an s-orbital centered at each atomic site and seek the conditions under which the lowest-energy multicentered bonding orbital $\psi(\vec{r})$, formed by the in-phase linear combination of the orbital on each site, has a local maximum at the center $\vec{r}_{\mathrm{c}}$ of the interstitial cage.
We do this by considering the first and second derivatives of the multicentered bonding orbital at the cage center, i.e., we evaluate $\nabla \psi(\vec{r}) |_{\vec{r} \to \vec{r}_{\mathrm{c}}}$ and the eigenvalues of the Hessian ($\mathrm{Eig}[\mathbf{H_\psi}(\vec r)|_{\vec r \to \vec r_c}]$).
For there to be a maximum, two criteria must be satisfied:

\begin{equation}
	\begin{aligned}
		\nabla \psi(\vec r) |_{\vec r \to \vec r_c} &= 0 \\
		\mathrm{Eig}[\mathbf{H_\psi}(\vec r)|_{\vec r \to \vec r_c}]  &< 0.
	\end{aligned}
\end{equation}

%Using regular polygons as examples, upper panel of Fig.~\ref{fig:2D-model} shows the schematics of our model where the center of the cage is set to be the origin of the Cartesian coordinate system and the cage size is controlled by the distance between the origin and corner atoms ($b$).
%

% \textcolor{blue}{[AAM: Be consistent in your suffix for denoting corner atoms, ie, $r_n$ vs $r_i$. The center of the $N$-gon / $N$-hedron is a vector, so most of the time when you write $r_c$ you actually mean $\vec{r}_{\mathrm{c}}$. The critical cage size should be $b_{\mathrm{c}}$ rather than $b_c$.]}
%
\subsection{Illustrative case with hydrogenic s-orbitals}
To illustrate our approach, we will present in detail the specific case of a 2D triangular cage (Fig.~\ref{fig:2D-model}, upper left panel), with a hydrogenic 1s-orbital, $\phi(r)=\frac{2}{\sqrt{4\pi}} e^{-r}$ in atomic units, on each corner atom. The multicentered bonding orbital is given by
\begin{equation}
	\begin{aligned}
		\psi(\vec r) = \sum_{n=1}^{3} \left[ \frac{2}{\sqrt{4\pi}} e^{-|\vec{r}-\vec{r}_n|}\right],
	\end{aligned}
\end{equation}
%\textcolor{blue}{[AAM: If we are neglecting normalisation, why not just get rid of the 2/sqrt 4pi?]}
%\textcolor{orange}{[CCX: 2/sqrt 4pi originates from the normalization of the atomic orbitals and was kept during the derivation so that in the normalization section (now deleted), the lower limit where I assume all overlap=1 exactly cancels out the $N$ dependencies in the Hessian eigenvalues.
%Also, if we delete it we also need to scale Fig. S2 accordingly...]}
where we have neglected normalization of the orbital as it does not affect its topology. Without loss of generality, we take the center of the triangle $\vec{r}_{\mathrm{c}}$ to be the origin. The Cartesian coordinates of the three corner atoms $\vec{r}_n$ are then given by
\begin{equation}
%	\begin{aligned}
		\vec r_1 = b\left(-\frac{\sqrt{3}}{2},-\frac{1}{2}\right), \quad
		\vec r_2 = b\left(\frac{\sqrt{3}}{2},-\frac{1}{2}\right) \quad \mbox{and}\quad
		\vec r_3 = b\left(0,1\right).
%	\end{aligned}
\end{equation}
It is straightforward to calculate $\nabla \psi = \left(\frac{\partial \psi}{\partial x}, \frac{\partial \psi}{\partial y}\right)$ and verify that it is zero at the cage center $\vec{r}_{\mathrm{c}}$.
%\begin{equation}
%\begin{aligned}
%	\frac{\partial\psi}{\partial x} &=  \frac{2}{\sqrt{4\pi}}
%	\frac{ \left(\frac{-\sqrt{3} b}{2}-x\right) e^{-\sqrt{\left(-\frac{\sqrt{3} b}{2}-x\right)^2+\left(-\frac{b}{2}-y\right)^2}}}{\sqrt{\left(\frac{-\sqrt{3} b}{2}-x\right)^2+\left(-\frac{b}{2}-y\right)^2}}
%	-
%	\frac{ x e^{-\sqrt{(b-y)^2+x^2}}}{\sqrt{(b-y)^2+x^2}}
%	+
%	\frac{ \left(\frac{\sqrt{3} b}{2}-x\right) e^{-\sqrt{\left(\frac{\sqrt{3} b}{2}-x\right)^2+\left(-\frac{b}{2}-y\right)^2}}}{\sqrt{\left(\frac{\sqrt{3} b}{2}-x\right)^2+\left(-\frac{b}{2}-y\right)^2}}\\
%	\frac{\partial\psi}{\partial y} &= \frac{2}{\sqrt{4\pi}}
%	\frac{ \left(\frac{-b}{2}-x\right) e^{-\sqrt{\left(-\frac{\sqrt{3} b}{2}-x\right)^2+\left(-\frac{b}{2}-y\right)^2}}}{\sqrt{\left(\frac{-\sqrt{3} b}{2}-x\right)^2+\left(-\frac{b}{2}-y\right)^2}}
%	+
%	\frac{ (b-y) e^{-\sqrt{(b-y)^2+x^2}}}{\sqrt{(b-y)^2+x^2}}
%	+
%	\frac{ \left(\frac{-b}{2}-x\right) e^{-\sqrt{\left(\frac{\sqrt{3} b}{2}-x\right)^2+\left(-\frac{b}{2}-y\right)^2}}}{\sqrt{\left(\frac{\sqrt{3} b}{2}-x\right)^2+\left(-\frac{b}{2}-%\end{aligned}
%\end{equation}
%One can easily verify that, at the cage center ($x=0,y=0$), the gradient of $\psi$ is 0, 
$\psi(\vec{r})$, therefore, has a turning point at $\vec{r}_{\mathrm{c}}$. The curvature at this point can be determined from the Hessian of $\psi$,
\begin{equation}
	\mathbf{H_\psi}(\vec r)|_{\vec r \to \vec r_c}=
	\frac{1}{\sqrt{4\pi}} \frac{ 3  \left(b-1\right) e^{-b}}{b}
    \begin{pmatrix}
	1 & 0\\
	0 & 1
	\end{pmatrix},
	\label{eq:2D-hes}
\end{equation}
%\textcolor{blue}{[AAM: shall we remove the 2/sqrt 4pi?]}
%\textcolor{orange}{[CCX: see above]}
which is isotropic, as expected by symmetry.
%From \ref{eq:2D-hes}, we see that the the Hessian matrix $\mathbf{H_{\psi}}$ is diagonal and the eigenvalues are degenerate. 
%This is caused by the 3-fold rotation symmetry of the model, and it means that the second order derivative at the origin is unidirectional.
%constraining the Hessian ellipse to be a circle.
% apply three fold rotation to the Hessian ellipse?
%
%Taking a closer look at the diagonal components of the Hessian, we see that the sign of t
It can be seen that the eigenvalues of the Hessian $\mathbf{H_\psi}$ depend solely on the cage size $b$.
%As shown in the lower left panel of Fig.~\ref{fig:2D-model}, for 1s-orbital, there is only one critical point (denoted as $b_c$) where $\text{Eig}(\mathbf{H_\psi})=0$ at 1~Bohr.
We denote $b^{\mathrm{H[2D]}}_{\mathrm{c}}=1~\mathrm{bohr}$ as the critical cage size for this triangular model with hydrogenic 1s-orbitals: when $b>b^{\mathrm{H[2D]}}_{\mathrm{c}}$, the eigenvalues are positive and $\psi$ has a minimum at $\vec{r}_{\mathrm{c}}$; when $b<b^{\mathrm{H[2D]}}_{\mathrm{c}}$, the eigenvalues are negative and $\psi$ has a (local) maximum at $\vec{r}_{\mathrm{c}}$ and, in other words, $\psi$ is a multicentered bonding orbital localized in the interstitial region.

\begin{figure}[H]
     \centering
     \includegraphics[scale=0.9]{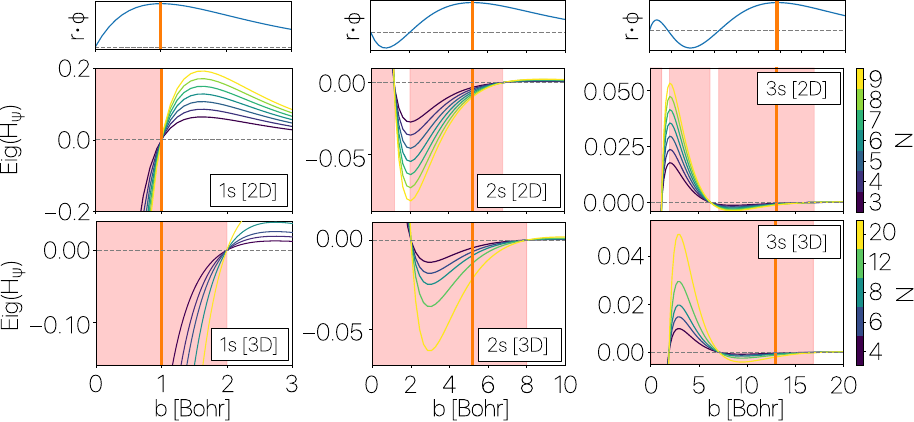}
     \caption{
     The radial distribution functions (RDFs), $r\phi(r)$, of the atomic orbitals and the eigenvalues of the Hessian at $\vec{r}_{\mathrm{c}}$ for regular polygons (upper panel) and 3D polyhedrons (lower panel) with different number of corners $N$. The red shaded regions represent the values of $b$ where the system has a maximum at $\vec{r}_{\mathrm{c}}$. The orange vertical lines highlight the position of the outermost maximum of the RDF for each $n$s-orbital, which we take as our measure of the size $r_0$ of the corresponding orbital. 
     % \textcolor{blue}{[AAM: there are also orange vertical lines at the nodes of the RDFs in the lower panel. Remove these or describe them in the caption?[done] The RDFs are the same in both panels. Can we combine and just have 3$\times$3 sub-panels?[done]]}
     }
     % Upper panel: schematics of the 3D regular polyhedron model where the origin point ($r_c$) is the geometric center of the polyhedron. The distance between $r_c$ and position of corner atoms ($r_n$) is denoted as $b$. Lower panel: radial distribution function ($r \cdot \psi$) of 1s, 2s and 3s orbitals and the corresponding eigenvalues of the Hessians at $r_c$ of the polyhedrons with different number of corners $N$. The red regions represent the values of $b$ where the system has a maxima at $r_c$. The Orange lines show the radius of the corresponding s-orbitals (defined as the radius of the outermost maxima point in the radial distribution function).}
     \label{fig:3D-model}
\end{figure}

The analysis above can be generalized to arbitrary regular polygons (in 2D) and polyhedrons (in 3D), and to hydrogenic $n$s-orbitals with $n>1$.  
Fig.~\ref{fig:3D-model} shows the dependency of Hessian eigenvalues at $\vec{r}_{\mathrm{c}}$ on the cage size $b$ for 2D polygons and 3D polyhedral cages with different numbers of corners $N$ and with 1s, 2s and 3s-orbitals on the corner atoms.
The red shaded regions in Fig.~\ref{fig:3D-model} represent the values of $b$ where the Hessian eigenvalue is negative and the multicentered bonding orbital has a local maximum at the cage center.
As can be seen, for both 2D and 3D cages, the critical cage size $b^\mathrm{H}_\mathrm{c}$ (for 2s and 3s-orbitals, $b^\mathrm{H}_\mathrm{c}$ is defined as the outer-most turning point of the Hessian eigenvalue) increases alongside the the principle quantum numbers of the atomic s-orbitals. 
An interesting feature of the model is that $b^{\mathrm{H}}_{\mathrm{c}}$ is independent of the number of corner atoms $N$. Moreover, $b^{\mathrm{H}}_{\mathrm{c}}$ is always equal to or greater than the corresponding s-orbital size $r_0$ (marked by the vertical orange lines in Fig.~\ref{fig:3D-model}). 
Table~\ref{tab:RDF-hes-formula} lists analytical formula for the Hessian eigenvalues at $\vec{r}_{\mathrm{c}}$ and the corresponding $b^\mathrm{H}_\mathrm{c}$ for 2D and 3D models with $n$s-orbitals for $n=1$ to $6$.

\begin{table}[!ht]
    \centering
    \caption{Critical cage sizes $b^\mathrm{H}_{\mathrm{c}}$ and eigenvalues of the Hessian of the multicentered bonding orbital $\phi(\vec{r})$ at the cage center $\vec{r}_{\mathrm{c}}$ formed by $n$s-orbitals on regular polygons (2D) and polyhedrons (3D). $b^\mathrm{H}_{\mathrm{c}}$ is independent of the number $N$ corner atoms of the regular 2D polygon or 3D polyhedron.}
    \begin{ruledtabular}
    \begin{tabular}{cccc}
 Dimension & $n$ & $b^\mathrm{H}_{\mathrm{c}}$ [\AA] & $\mathrm{Eig}[\mathbf{H_\psi}(\vec{r})|_{\vec{r} \to \vec{r}_{\mathrm{c}}}]$ [a.u.] \\ \hline
\multirow{6}{*}[-3em]{2D}  & 1 & 0.529  & $\frac{\sqrt{\frac{1}{4 \pi }} \left[(b-1) e^{-b}\right] N}{b}$ \\
                           & 2 & 3.61  & $-\frac{\sqrt{\frac{1}{4 \pi }} \left[\left(-\frac{b^2}{8}+b-1\right) e^{-\frac{b}{2}}\right] N}{\left(2 \sqrt{2}\right) b}$\\
                           & 3 & 8.77  & $\frac{\sqrt{\frac{1}{4 \pi }} \left[\left(\frac{2 b^3}{81}-\frac{16 b^2}{27}+\frac{29 b}{9}-3\right) e^{-\frac{b}{3}}\right] N}{\left(9 \sqrt{3}\right) b}$\\
                           & 4 & 16.0  & $-\frac{\sqrt{\frac{1}{4 \pi }} \left[\left(-\frac{b^4}{256}+\frac{13 b^3}{64}-3 b^2+\frac{27 b}{2}-12\right) e^{-\frac{b}{4}}\right] N}{96 b}$\\
                           & 5 & 25.4  & $\frac{\sqrt{\frac{1}{4 \pi }} \left[\left(\frac{8 b^5}{15625}-\frac{152 b^4}{3125}+\frac{928 b^3}{625}-\frac{432 b^2}{25}+\frac{348 b}{5}-60\right) e^{-\frac{b}{5}}\right] N}{\left(300 \sqrt{5}\right) b}$\\
                           & 6 & 37.0  & $-\frac{\sqrt{\frac{1}{4 \pi }} \left[\left(-\frac{b^6}{17496}+\frac{13 b^5}{1458}-\frac{235 b^4}{486}+\frac{305 b^3}{27}-\frac{1025 b^2}{9}+\frac{1280 b}{3}-360\right) e^{-\frac{b}{6}}\right] N}{\left(2160 \sqrt{6}\right) b}$\\ \hline
 \multirow{6}{*}[-3em]{3D} & 1 &  1.06 & $\frac{\sqrt{\frac{1}{4 \pi }} \left[\left(\frac{2b}{3}-\frac{4}{3}\right) e^{-b}\right] N}{b}$ \\
                           & 2 &  4.23  & $-\frac{\sqrt{\frac{1}{4 \pi }} \left[\left(-\frac{b^2}{12}+\frac{5 b}{6}-\frac{4}{3}\right) e^{-\frac{b}{2}}\right] N}{\left(2 \sqrt{2}\right) b}$\\
                           & 3 &  9.53  & $\frac{\sqrt{\frac{1}{4 \pi }} \left[\left(\frac{4 b^3}{243}-\frac{4 b^2}{9}+\frac{26 b}{9}-4\right) e^{-\frac{b}{3}}\right] N}{\left(9 \sqrt{3}\right) b}$\\
                           & 4 &  16.9  & $-\frac{\sqrt{\frac{1}{4 \pi }} \left[\left(-\frac{b^4}{384}+\frac{7 b^3}{48}-\frac{19 b^2}{8}+\frac{25 b}{2}-16\right) e^{-\frac{b}{4}}\right] N}{96 b}$\\
                           & 5 &  26.5  & $\frac{\sqrt{\frac{1}{4 \pi }} \left[\left(\frac{16 b^5}{46875}-\frac{64 b^4}{1875}+\frac{416 b^3}{375}-\frac{352 b^2}{25}+\frac{328 b}{5}-80\right) e^{-\frac{b}{5}}\right] N}{\left(300 \sqrt{5}\right) b}$\\
                           & 6 &  38.1  & $-\frac{\sqrt{\frac{1}{4 \pi }} \left[\left(-\frac{b^6}{26244}+\frac{b^5}{162}-\frac{85 b^4}{243}+\frac{700 b^3}{81}-\frac{850 b^2}{9}+\frac{1220 b}{3}-480\right) e^{-\frac{b}{6}}\right] N}{\left(2160 \sqrt{6}\right) b}$\\
    \end{tabular}
    \end{ruledtabular}
    \label{tab:RDF-hes-formula}
\end{table}

\subsection{More realistic model with DFT-calculated atomic orbitals}
The analytical results presented in the previous section with hydrogenic $n$s-orbitals provide informative, qualitative insight. For more quantitative results that are more reflective of the behaviour of real materials, however, it is better to use the relevant atomic radial orbitals calculated from density functional theory, rather than analytical hydrogenic s-orbitals.
%always more disperse than the real atomic radial wavefunctions due to the imperfect screening of the core charges.
%To get a set of $b_c$ that's better suited for the real situations, we need to modify the form of our atomic orbitals.
As we saw above, the critical cage size is always larger than the size of the orbital $r_0$ and, therefore, it is the overlap of the tails of orbitals that is responsible for the local maximum of the multicentered bonding orbital $\psi(\vec{r})$ at the center of the cage $\vec{r}_{\mathrm{c}}$. For this reason, rather than using numerical radial orbitals directly, we fit their tails (i.e., for $r>r_0$) to an analytic form given by
%One way to obtain a more realistic critical cage size is to fit the tails of the real radial wavefunctions to a pseudo radial wavefunctions $\mathrm{WF_{fit}}(r')$:
\begin{equation}
%	\mathrm{WF_{fit}}(r') = D  Z^{3/2} \text{WF}[Z \cdot (r'+C)].
	\tilde{\phi}_{n}(r) = D  Z^{3/2} \phi_{n}(Z(r+C)),
	\label{eq:WF_fit}
\end{equation}
where $\phi_n(r)$ is a hydrogenic $n\mathrm{s}$-orbital (e.g., $\phi_1(r)=\frac{2}{\sqrt{4\pi}} e^{-r}$, $\phi_2(r)=\frac{\sqrt{2}}{8\sqrt{\pi}}(2-r)e^{-r/2}$, etc.), and $C$, $D$ and $Z$ are fitting parameters. The multicentered bonding orbital is then written as a linear combination of these fitted functions for $r>r_0$,
\begin{equation}
\psi(\vec{r}) = \sum_{i=1}^{N} \tilde{\phi}_{n}(|\vec{r}-\vec{r}_i|),
\end{equation}
where the atoms are at the $N$ corners $\{r_i\}$ of the cage.

\begin{figure}[H]
     \centering
     \includegraphics[scale=0.6666]{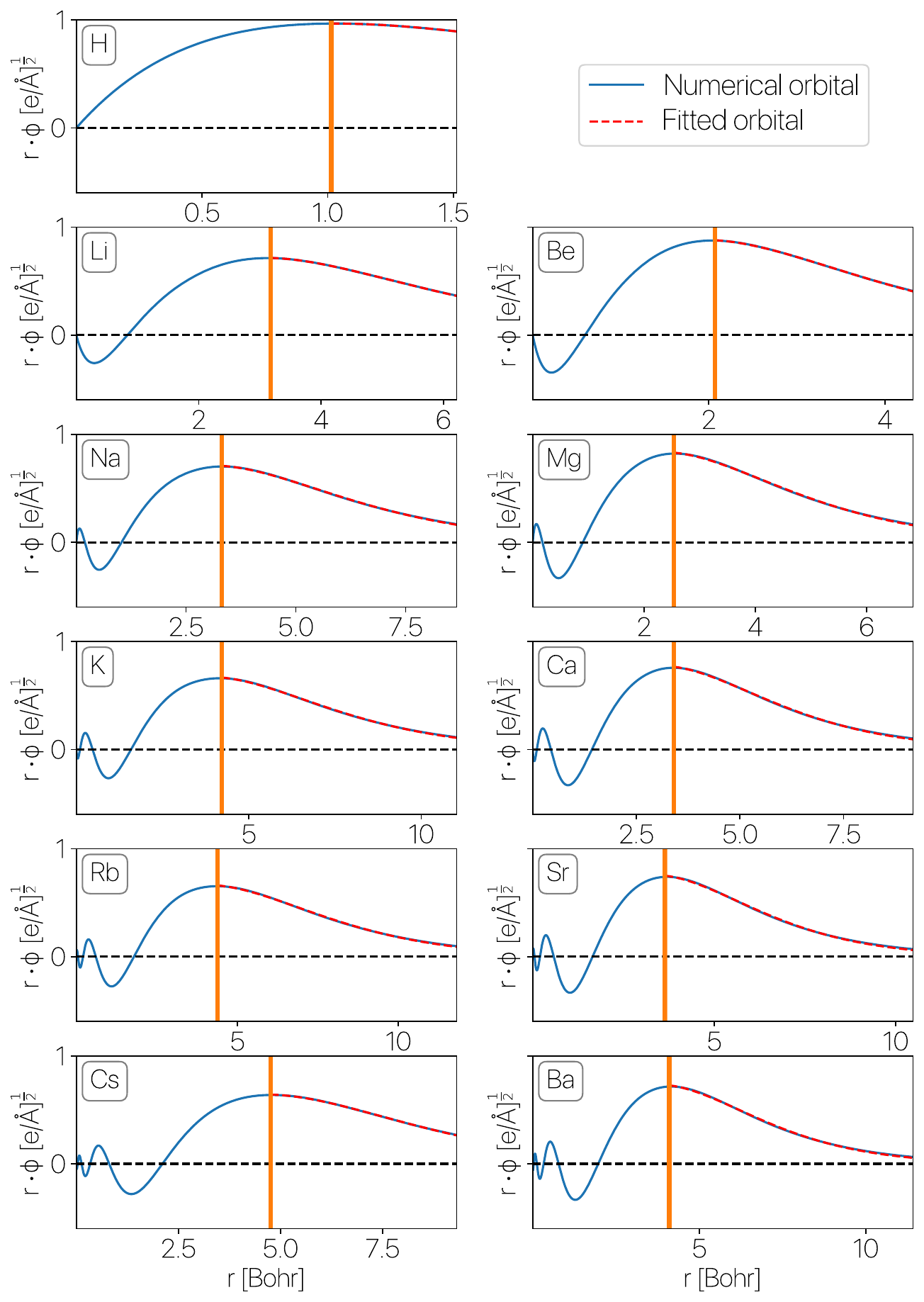}
     \caption{RDF [$r\phi(r)$] of the frontier s-orbitals (blue curves) of Group I and II elements (up to the 6th period), obtained from VASP's PAW dataset (PBE version 54) and the analytical fit (Eq.~\ref{eq:WF_fit}) to their tails (red dashed curves). Orange vertical lines indicate the position $r_0$ of the maximum of the RDF.}
     % \textcolor{blue}{[AAM: the legend says `AE partial wave', which is a term that is never mentioned anywhere. Maybe replace with `Numerical orbital', and `Fitted orbital' underneath it?]}}
     \label{fig:element-RDF}
\end{figure}

Fig.~\ref{fig:element-RDF} shows the results of the fitting for the frontier s-orbitals of group I and II elements. It is clear that the tails of the radial distribution functions (RDFs) are reproduced to high accuracy.
Because of the linear relation between $\tilde{\phi}_n(r)$ and $\phi_n(r)$ in Eq.~\ref{eq:WF_fit}, we can use the analytical results of the previous section to obtain a set of new critical values $b_{\mathrm{c}}$ that correspond to this more realistic choice for the atomic s-orbitals, and it is straightforward to show that
\begin{equation}
	b_{\mathrm{c}} = \frac{b^{\mathrm{H}}_{\mathrm{c}}}{Z}-C.
\end{equation}
The fitted parameters and resulting critical values $b_{\mathrm{c}}$ are given in Table~\ref{tab:critical_b}. These are the values that we use in this work and that are given in main manuscript.

\begin{table}[!ht]
    \centering
    \caption{Fitting parameters $Z$, $D$, $C$ and resulting critical cage sizes $b_{\mathrm{c}}$ of the multicentered bonding orbital $\phi(\vec{r})$ at the cage center $\vec{r}_{\mathrm{c}}$ formed by s-orbitals of group I and II elements on regular polygons (2D) and polyhedrons (3D).}
    % \textcolor{blue}{[AAM: Can you please check carefully that the numbers below match exactly the numbers in the Table in the main text?[done] Also, please can you align all these numbers at the decimal point?[done]]}}
    % \textcolor{blue}{[AAM: I would keep the numbers to a consistent 3 significant figures (rather than 3 decimal places). As noted in another comment, I think you can move the hydrogenic results to the other table.]}}
    \begin{ruledtabular}
    \begin{tabular}{ccc *{5}{d{3.3}}}
    % \begin{tabular}{S[table-format=3.2]}
    % \begin{tabular}{S[table-format=3.2]}
\mc{Group} & \mc{Element} & \mc{$n$} & \mc{$Z$} & \mc{$D$} & \mc{$C$ [\AA]} & \mc{$b^\mathrm{[3D]}_{c}$ [\AA]} & \mc{$b^\mathrm{[2D]}_{c}$ [\AA]}\\ \hline
\multirow{6}{*}{I}&H  & 1   &   0.922   &  1.21  &  0.321    & 1.12  &  0.542   \\
                  &Li & 2	&   1.34   &  2.20  &  0.444   & 2.71  & 2.25  \\
                  &Na & 3	&   2.13   &  4.75  &  1.59    & 2.88  & 2.52  \\
                  &K	& 4 &   2.77   &  7.33  &  2.65    & 3.47  & 3.15  \\
                  &Rb	& 5 &   3.61   &  11.3  &  3.70    & 3.63  & 3.35  \\
                  &Cs	& 6 &   4.63   &  16.2  &  4.29    & 3.94  & 3.70  \\
\hline
\multirow{5}{*}{II}&Be	& 2 &   1.94   &  3.92  &  0.384  & 1.80  & 1.48  \\
                   &Mg	& 3 &   2.85   &  7.47  &  1.18   & 2.17  & 1.90  \\
                   &Ca	& 4 &   3.59   &  10.9  &  1.97   & 2.74  & 2.50  \\
                   &Sr	& 5 &   4.60   &  16.3  &  2.80   & 2.95  & 2.73  \\
                   &Ba	& 6 &   5.52   &  22.0  &  3.63   & 3.27  & 3.07  \\
    \end{tabular}
    \end{ruledtabular}
    \label{tab:critical_b}
\end{table}

\clearpage 
%\section{\label{sec:RDF} RDF of frontier s-orbitals of Group I and II elements}

% \textcolor{blue}{[AAM: In this section, we denote the multicentered bonding orbital by $\phi$. But when we mention the RDF of the atomic orbitals, we also call this $r\phi(r)$ in the text, the figures and the captions. This is potentially confusing for a reader. In one of the parts that I've added, I've called the atomic orbitals $\phi_n(r)$ or $\tilde{\phi}_n(r)$, so maybe we can use $r\phi(r)$ for RDF?]}

\clearpage 
\section{Calculation details}
\label{sec:Methods}

The electronic structures of the materials studied in Sec.~\ref{sec:validation} and  \ref{sec:results} in the main manuscript and Sec.~\ref{sec:bonding_analysis}, \ref{sec:organic_electrides} and \ref{subsec:other_electrides;F-center} in the supplemental material were evaluated using first-principles calculations carried out within the density functional theory formalism, as implemented in the Vienna Ab initio Simulation Package (VASP) code \citeS{SKresse1996b,SKresse1993,SKresse1996a}.
The Materials Project (MP) database provides all the necessary inputs optimized by a series of meticulously designed protocols~\citeS{SJain2013,SJain2011}. 
Taking advantage of these previously validated inputs, we have re-calculated the electronic structures while enabling the output of the ELFs and all-electron charge densities.
% for the 51,913 structures. 

More specifically, in our DFT calculations, the Perdew-Burke-Ernzerhof (PBE) variant of the generalized gradient approximation (GGA)~\citeS{SBlochl1994} was employed as the exchange-correlation functional.
The projector augmented wave (PAW) method was used to model the valence wavefunctions within the core regions~\citeS{SKresse1999} while the frozen core approximation was employed for core electrons.
The valence wavefunctions were expanded using a plane-wave basis set with a kinetic-energy cutoff of 520 eV, as employed by the MP database (except for the high-pressure hP4-Na where a kinetic energy cutoff of 910~eV was adopted).
To sample the first Brillouin zone of the unit cell, a uniform Monkhorst-Pack mesh \citeS{SMonkhorst1976} with a total of 1000/(number of atoms in the unit cell) k-points was employed and the tetrahedron method (for predicted insulators) or Gaussian smearing with a width of 0.5~eV (for predicted metals) were used to perform the k-point integration.
The self-consistent field calculations were carried out with an energy convergence criteria of $\mathrm{10^{-5}}$ eV for the high-throughput calculations and a stricter $\mathrm{10^{-8}}$ eV was used for the structures mentioned in the main manuscript (hP4-Na, Ca$_2$N and Hf$_2$Au).
To alleviate the self-interaction error introduced by the exchange-correlation functional, DFT+U was employed to some systems that contain: Co, Cr, Fe, Mn, Mo, Ni, V and W, where the choice of the U values were obtained by fitting to experimental binary formation enthalpies as described in Wang et al \citeS{SWang2006,SJain2011prb}, as employed by the MP database.
For the predicted magnetic systems, the magnetic order and corresponding magnetic moments reported by the MP database were used as the input for the DFT calculations.
All structures used in the high-throughput studies were taken directly from the Materials Project without further relaxation. 
For hP4-Na, Ca$_2$N and Hf$_2$Au the atomic positions and the unit cells were relaxed until the forces between atoms are smaller than 0.001eV/\AA. 
The atomic coordinates and all other structural information can be found in Table.~\ref{tab:structure_params}.

%All other parameters such as the energy cutoff, k-point sampling and DFT+U corrections were taken directly from the Materials Project without any modification.
%The valence wavefunctions were expanded using a plane-wave basis set with a kinetic-energy cutoff of 520 eV, as employed by the MP database (except for the high-pressure hP4-Na where a cutoff of 910~eV was adopted).
%To sample the first Brillouin zone of the unit cell, a uniform Monkhorst-Pack mesh \citeS{SMonkhorst1976} with a total of 1000/(number of atoms in the unit cell) k-points was employed and the tetrahedron method (for predicted insulators) or Gaussian smearing with a width of 0.5~eV (for predicted metals) were used to perform the k-point integration.
%To levitate the self-interaction error introduced by the exchange-correlation functional, DFT+U was employed to some systems that contain: Co Cr, Fe, Mn, Mo, Ni, V and W, where the choice of the U values were obtained by fitting to experimental binary formation enthalpies as described in Wang et al \citeS{SWang2006,SJain2011prb}, as employed by the MP database.

%All calculations were performed without considering spin polarization except for the systems that were identified to have a stable magnetic order by the MP database.
%For the predicted magnetic systems, the magnetic order and corresponding magnetic moments reported by the MP database were used as the input for the DFT calculations.

All calculation inputs can be found \href{https://zenodo.org/records/7474009?token=eyJhbGciOiJIUzUxMiIsImlhdCI6MTc3NzM3OTYwOCwiZXhwIjoxNzg1NTQyMzk5fQ.eyJpZCI6ImNjNmU2ZWUyLTBlODQtNDc0NC1hMTYwLWI0MzA0ZGY3OWYwNiIsImRhdGEiOnt9LCJyYW5kb20iOiJkZjU5OTk4MjE0ZTA0NDgxOWE4Yjc3ZDgxM2RjN2VkNCJ9.X4kBZBNP8_f6Gkc4uMKnu1HRDY8DLbJ_DoXYYTb3n4g_H-5u5IpgXCbJ7I-S7yYJ198kNVflN0LTt4DH2uw13g}{Zenodo}. 
%A searchable database can be found at \href{https://contribs.materialsproject.org/projects/electride_data_base_cxiao}{the MPcontrib page}.

\begin{table}[!ht]
    \centering
    \caption{Calculated lattice parameters and atomic positions for Ca$_2$N, hP4-Na and Hf$_2$Au in Cartesian coordinates.}
    \begin{ruledtabular}
    \begin{tabular}{ccc*{3}{d{3.3}}}
Formula &   &  & \mc{x [\AA]} & \mc{y [\AA]} & \mc{z [\AA]} \\ \hline
\multirow{6}{*}[0em]{Ca$_2$N} & \multirow{3}{*}[0em]{Lattice parameters} & a & 6.379 & -0.106 & -0.064 \\
& & b & 5.410 & 3.380 &  -0.064 \\
& & c & 5.410 & 1.503 &  3.029 \\ \cline{2-6}
& \multirow{3}{*}[0em]{Atomic positions} & Ca & 12.655 & 3.515 & 2.134 \\
& & Ca & 4.545 & 1.262 & 0.766 \\ 
& & N & 0.000 & 0.000 & 0.000 \\ \hline

\multirow{6}{*}[-0.5em]{hP4-Na} & \multirow{3}{*}[0em]{Lattice parameters} & a & 2.786 & 0.000 & 0.000 \\
& & b & 1.393 & 2.412 & 0.000 \\
& & c & 0.000 & 0.000 &  3.867 \\ \cline{2-6}
& \multirow{3}{*}[-0.5em]{Atomic positions} & Na & 0.000 & 0.000 & 0.000 \\
& & Na & 1.393 & 0.804 & 0.967 \\ 
& & Na & 0.000 & 0.000 & 0.000 \\ 
& & Na & 2.786 & 1.608 & 2.900 \\ \hline

\multirow{6}{*}[0em]{Hf$_2$Au} & \multirow{3}{*}[0em]{Lattice parameters} & a & 6.278 &  -0.013 &  0.000 \\
& & b & 4.606  &  4.266  &  0.000 \\
& & c & -5.442 &  -2.126  &  2.297 \\ \cline{2-6}
& \multirow{3}{*}[0em]{Atomic positions} & Au & 0.000 & 0.000 & 0.000 \\
& & Hf & 7.205 & 2.815 & 0.000 \\ 
& & Hf & 3.679 & 1.437 & 0.000 \\

    \end{tabular}
    \end{ruledtabular}
    \label{tab:structure_params}
\end{table}

\newpage
\section{Projection using PAW pseudo-partial waves vs using PAW projectors}
\label{sec:proj_PAW_partial_wave}
Projected band structure is a powerful tool that can be used to demonstrate the atomic contributions to the electronic band.
By projecting the DFT calculated wavefunctions ($\Psi_{n,\vec k}$) onto one(or multiple) atomic orbitals ($p^{\alpha}_{lm}$), the projection coefficient can be written as,
\begin{equation}
	P^{\alpha}_{lm,n\vec k } = \left|\braket{p^{\alpha}_{lm}|\Psi_{n,\vec k}}\right|^2,
\end{equation}
where $\alpha$ is the site index of the atomic orbital and $l$,$m$ are its angular quantum numbers. $n$ and $\vec k$ represent the band number and k-point of the DFT wavefunction.
The projection function $p^{\alpha}_{lm}$ is usually constructed with a radial and an angular part.
Whilst the angular part of $p^{\alpha}_{lm}$ is simply real spherical harmonics, the radial part (which determines the size of the atomic orbital) has multiple choices in standard DFT codes.

VASP uses the projector augmented wave method~\citeS{SKresse1999} where the variational quantity, the pseudo wavefunction $\tilde \Psi$, differs from the true Kohn-Sham wavefunction (or all-electron wavefunction, $\Psi$) only inside an atom centered sphere with a PAW cutoff radius $r_\mathrm{PAW}$. 
The pseudo wavefunctions can be transformed to the all-electron wavefunctions by:
\begin{equation}
\ket{\Psi}=\left[1+\sum_i\left(\ket{\phi_i}-\ket{\tilde{\phi}_i}\right)\bra{p_i}\right]\ket{\tilde{\Psi}}
\end{equation}
where ${\phi}_i$ represents a set of pre-calculated all-electron atomic wavefunctions (all-electron partial waves) and their pseudo counterpart $\tilde{\phi}_i$ (pseudo partial waves). 
$p_i$ are the so-called PAW projectors that are dual to the $\tilde{\phi}_i$ and exist strictly inside the PAW sphere. i.e.,
\begin{equation}
    \braket{p_i|\tilde{\phi}_j} = \int_{0}^{r_\mathrm{PAW}} p_i(r) \tilde{\phi}_j(r) dr = \delta_{ij}.
\end{equation}

In VASP, the radial part of the projection functions are usually chosen as the PAW projectors (\verb|IORBIT=11|, VASP version 5.4.4pl2).
However, because the PAW projector is confined inside the PAW sphere, projection coefficients calculated with it only represent the atomic features of the DFT wavefunctions inside the PAW sphere.
As such, the projection coefficients calculated using the PAW projectors are usually weighted by the fraction of the all-electron charge inside the PAW sphere $Q_\textrm{tot}$, defined as:
\begin{equation}
	Q_\textrm{tot} = \int_0^{r_{\mathrm{PAW}}} \left | r \phi(r)\right |^2 dr
	\label{eq:qtot}
\end{equation}
where the quantity $r \phi(r)$ can be read from the PAW dataset.
%From Eq.~\ref{eq:qtot}, we see that $Q_{tot}$ is the fraction of the all-electron charge inside the PAW cutoff sphere so the final projection coefficient 
% 
%Note that the square of $r \cdot \psi_\mathrm{AE}$ is normalized to 1.
% Since in Eq. \ref{eq:qtot}, $r \psi_\mathrm{AE}$ is integrated from $0$ to the PAW cutoff radius ($r_\mathrm{PAW}$), $Q_{tot}$ is the fraction of the all-electron charges that are inside the PAW sphere.
% Based on this, the calculated projection coefficients 
%should be interpreted as the all-electron charge inside the PAW sphere that has the specified atomic features.
As a result of this implementation, atomic orbitals that are very disperse will have a very small $Q_\textrm{tot}$, effectively quenching the final projection values.

\begin{figure}[H]
     \centering
     \includegraphics[scale=1]{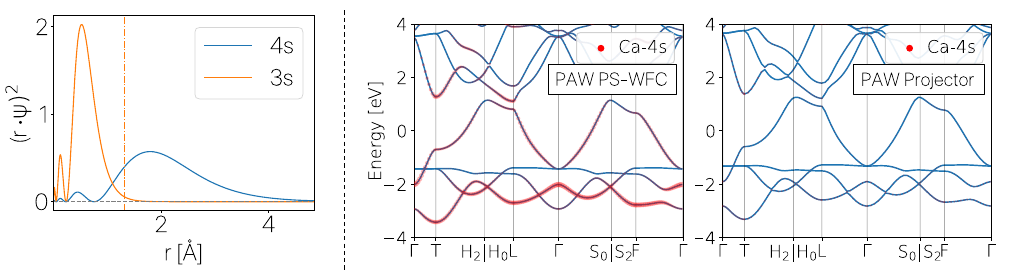}
     \caption{Left panel: radial distribution of the square of the radial distribution of Ca's 3s and 4s all-electron radial partial waves (PS-WFC). Dotted orange line indicates the PAW cutoff radius ($r_\mathrm{PAW}$); Right panel: bandstructures of the Ca$_2$N system projected onto Ca's 4-orbital using the PAW pseudo partial waves and the PAW projector. The size of the red circles indicate the size of the projection coefficients.}
     \label{fig:qtot}
\end{figure}

For example, Ca's PAW dataset has two projectors, representing the 3s and 4s states (see Fig.~\ref{fig:qtot}). 
While the radial distribution function of the 3s-orbital is almost fully enclosed by the PAW cutoff sphere, the 4s-orbital exists largely outside the PAW cutoff sphere.
As such, for an isolated Ca atom, the projection coefficient obtained for Ca's 3s is 0.987 while 4s only has a projection coefficient of 0.105.
The calculated $Q_\mathrm{tot}$ of Ca's 4s-orbital using its associated all-electron partial wave is 0.127 which is much smaller than the ideal value of 1, meaning the calculate projection coefficient ($\braket{p_i|\tilde{\phi}_j}$) only captures a small portion of the entire orbital. If we use the  full projector, the projection coefficient $P^{\alpha}_{lm,n\vec k}$ would be much higher.

%(slightly smaller than 1, caused by the periodic boundary conditions and simulated box size).
Projection functions built directly using the PAW pseudo partial waves, on the other hand, don't suffer from this.
With the PAW pseudo partial waves as the radial part of the projection operator (using \verb|LOCPROJ|, VASP version 5.4.4pl2), we obtained a projection coefficient of 0.999 for Ca's 4s state.

%We suspect this $Q_{tot}$ prefactor could be the culprit behind identifying the interstitial state to be an ``n" that cannot be described using the multicentered bonding theory.
%  [Don't want to be too harsh]
% , as suggested in the literature [REFs].
As shown in the right panel of Fig.~\ref{fig:qtot}, the PAW projector projected bandstructure of $\mathrm{Ca_2N}$ system shows that the Ca's 4s-orbital has negligible contribution to the relevant valence bands, leading to the false conclusion that these bands should be described as ``non-atomic'' states.
However, with the PAW pseudo partial waves, we can clearly see that the relevant valence manifold should be attributed to Ca's 4s-orbitals, indicating the multicentered-bonding nature of these interstitial localized states.

\newpage 
\section{Additional bonding analysis}
\label{sec:bonding_analysis}
\subsection{High pressure electrides hP4-Na} \label{subsec:hP4-Na}

\begin{figure}[H]
     \centering
     \includegraphics[scale=0.9]{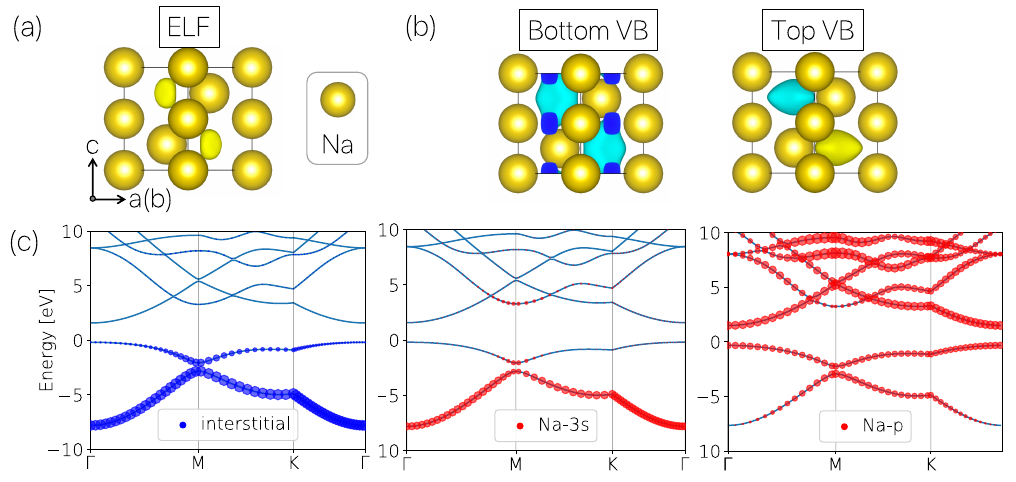}
     \caption{(a) ELF of the hP4-Na. (b) Wavefunction plots of the bottom valence band (around $-$8~eV to $-$2~eV) and the top valence band ($-$2~eV to 0~eV) at $\Gamma$ point. (c) Band structure projected onto: hydrogenic 1s-orbitals at the interstitial site (left panel); Na's s-orbital using PAW pseudo-partial waves (middle panel); and Na's p-orbitals using PAW projectors (right panel).}
%     (c) Band structure projected onto, left panel: hydrogenic 1s-orbitals at the interstitial site, middle panel: Na's s-orbital using PAW pseudo-partial waves and right panel: Na's p-orbitals using PAW projectors.}
     \label{fig:S1}
 \end{figure}

High pressure electrides provide a unique perspective into the origin of electrides and how they can be understood with the multicentered bonding theory.
For example, hP4-Na, stable under 320~GPa isotropic pressure has a double hexagonal closed-packed structure [Fig.~\ref{fig:S1}(a)] with a space group of P6$_3$/mmc.
There are two interstitial regions in the unit cell of hP4-Na which previous studies have shown to be filled with electrons~\citeS{SMa2009}.
As can be seen from Fig.~\ref{fig:S1}(a), in line with previous studies, our calculated ELF plot shows two interstitial peaks with ELF values around 0.98 and an integrated charge number of 1.98 on both sites.

The left panel of Fig.~\ref{fig:S1}(c) shows the bandstructure of hP4-Na projected onto two hydrogenic 1s-orbitals located at the identified interstitial site.
We can clearly see that both the top valence band and the bottom valence band  contribute significantly to the interstitial electronic states.
This finding is further confirmed by the wavefunctions plots of these two valence bands at $\Gamma$ point [see Fig.~\ref{fig:S1}(b)].
Instead of the in-phase combination of both interstitial states in the bottom valence band ($-$8~eV to $-$2~eV), the two interstitial pockets from the top valence band have opposite phase [see Fig.~\ref{fig:S1}(b)] and can largely be attributed to the multicentered bonding states between Na's 3p-orbitals (Fig.~\ref{fig:S1}(c), right panel).

\newpage
\subsection{Ca\texorpdfstring{\textsubscript{2}}{2}N type 2D electrides}
\begin{figure}[H]
     \centering
     \includegraphics[scale=1.0]{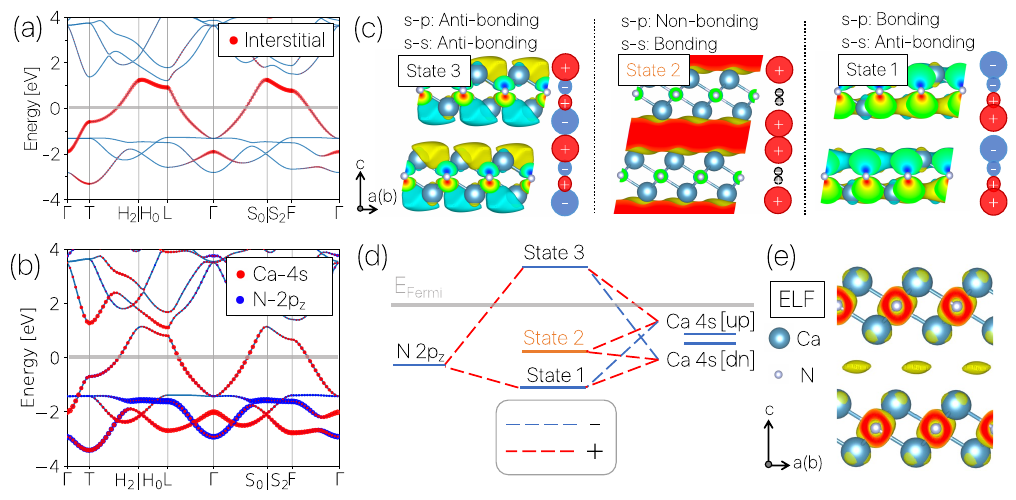}
     \caption{(a) Bandstructure plot of Ca$_2$N projected on to hydrogenic 1s-orbitals located at the identified interstitial sites. Fermi level is  adjusted to 0~eV, marked by the gray horizontal line. (b) Same as (a) but the projection was made onto Ca's 4s and N's 2p$_z$ orbitals using PAW pseudo partial waves. (c) Wavefunction plots of the relevant multicentered bonding states at $\Gamma$ point. (d) Bonding diagram of the Ca$_2$N involving Ca's 4s ([up]-upper layer; [dn]-lower layer) and N's 2p$_z$-orbitals. Red dotted lines means the atomic orbital has a positive contribution to the molecular orbitals and blue dotted lines means a negative contribution. 
     % \textcolor{red}{[CCX: change color of state 1 to blue, add `up' and `down' for Ca-4s]} 
     (e) ELF plot of Ca$_2$N.}
     \label{fig:Sup-Ca2N}
\end{figure}

%In section \ref{subsec:hP4-Na}, we've shown that 1D high pressure electride can be well described using the multicentered bonding theory.
%In this section, we will explore some more prototypical electride systems with the multicentered bonding theory.
%using the PAW pseudo partial wave projection method and wavefunctions plots.
2D A$_2$B electrides are prototypical electrides that have been proposed as superconductors and as potential hosts to the existence of Wigner crystals~\citeS{SGe2017,SKim2022}.
%https://iopscience.iop.org/article/10.1088/1367-2630/aa8a2d/pdf
%https://www.nature.com/articles/s41563-022-01353-8
Ca$_2$N as an example, has a crystal structure identical to that of the T-phase of transitional metal dichacognides where the interstitial sites are located at the centre of octahedrons in between Ca$_2$N layers [see Fig.~\ref{fig:Sup-Ca2N}(e)].
By projecting the DFT wavefunctions onto a set of hydrogenic 1s-orbitals at the interstitial sites, we identified the band that crosses the Fermi energy to be the main electride-like band that is responsible for the interstitial electronic behavior [Fig.~\ref{fig:Sup-Ca2N}(a)].
Our PAW pseudo partial wave projected bandstructure plot [Fig.~\ref{fig:Sup-Ca2N}(b)] suggests that this interstitial band between $-$2~eV and 1~eV [labeled as state 2 in Fig.~\ref{fig:Sup-Ca2N}(c,d)] is mainly composed of Ca's 4s-orbitals.
However, apart from state 2, Ca's 4s-orbitals also contribute to two other bands: the one between $-$3~eV and $-$2~eV [labeled as state 1 in Fig.~\ref{fig:Sup-Ca2N}(c,d)] and the one between 1.6~eV and 3.7~eV [labeled as state 3 in Fig.~\ref{fig:Sup-Ca2N}(c,d)].
% Ca's 4s-orbitals mainly contributes to two bands: the electride-like band , and the band between -3~eV and -2~eV [labeled as state 1 in Fig.~\ref{fig:Sup-Ca2N}(c,d)]. 

While the electride-like bands (state 2) solely comes from the bonding combination of Ca's 4s-orbitals between different Ca$_2$N layers, state 1 (state 3) can be considered the in-phase \emph{intralayer} bonding (anti-bonding) combination between Ca's 4s-orbitals and N's 4p$_z$-orbitals with an \emph{interlayer} anti-bonding combination between Ca atoms' 4s-orbitals.
% More specifically, the lowest energy bonding orbital comes from the bonding combination between two Ca atoms' 4s-orbitals and N's 4p$_z$-orbitals while the band that crosses the Fermi energy comes from the bonding stsate

\newpage
\subsection{Sr\texorpdfstring{\textsubscript{3}}{3}CrN\texorpdfstring{\textsubscript{3}}{3}}
\begin{figure}[H]
     \centering
     \includegraphics[scale=0.8]{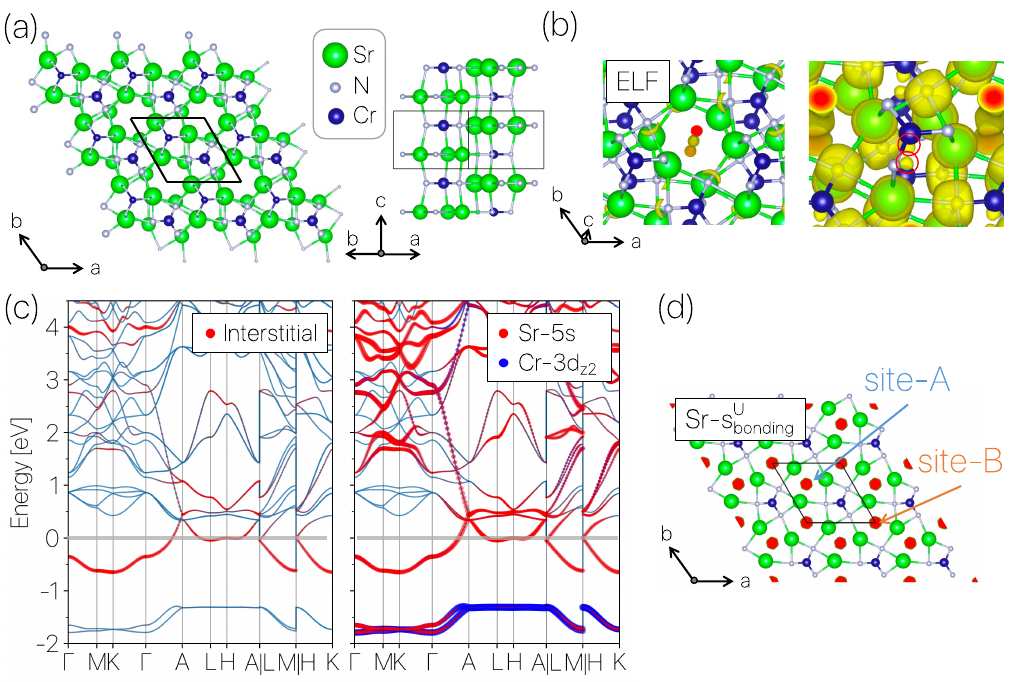}
     \caption{(a) Structure of Sr$_3$CrN$_3$, atoms in the upper layer are marked with a larger radius while atoms in the lower layer are maked with a smaller atomic radius. (b) Left panel: ELF plot indicating the existence of interstitial maxima. Right panel: two nearby maxima above and below Cr atoms.
     % formed by Cr's d$_{z^2}$-orbitals hybridizing with Sr's 5s-orbitals. 
     (c) Bandstructure projected on to hydrogenic 1s-orbitals at the interstitial sites, Sr's 5s-orbitals and Cr's 3d$_{z^2}$ orbitals using PAW projectors. (d) Structure illustration of the two interstitial bonding sites within the upper atomic layer (site-A and site-B).}
     \label{fig:Sr3CrN3}
\end{figure}

In prototypical 1D electride Sr$_3$CrN$_3$, electrons are identified to be localized in 1D channels of the crystal lattice~\citeS{SChanhom2019}.
The structure of Sr$_3$CrN$_3$ consists of two atomic layers, each with three Sr atoms, three N atoms and one Cr atom [Fig.~\ref{fig:Sr3CrN3}(a)]. 
These two layers are associated with each other with a 6-fold rotation and a translation operation, resulting in a space group of P6$_3$/m.

As shown in the left panel of Fig.~\ref{fig:Sr3CrN3}(b), the ELF indicates that there are two interstitial sites located in a 1D channel surrounded by Sr atoms, and both of these two interstitial sites have three nearest neighbor Sr atoms.
%By projecting the band structure onto two 1s-orbitals located at these two interstitial sites, 
The bandstructure projected onto hydrogenic 1s-orbitals located at these identified sites [Fig.~\ref{fig:Sr3CrN3}(c)] suggests that there are two relevant bands related to these interstitial sites, one at around $-$0.5~eV and the other one at around 4~eV.
Comparing to the bandstructure projected onto Sr's 5s and Cr's 3d$_{z^2}$-orbitals [Fig.~\ref{fig:Sr3CrN3}(c), right panel], we see that the interstitial band at around $-$0.5~eV mainly comes from Sr's 5s-orbital.
Besides the interstitial band, Sr's 5s-orbital also contributes to the two other bands, one at around $-$1.8~eV, hybridized with with Cr's 3d$_{z^2}$-orbitals [Fig.~\ref{fig:Sr3CrN3}(c), right panel] and the other one at around 4~eV.
% It turns out that these four bands are multicentered bonding states formed by Sr's 5s-orbitals and Cr's 3d$_{z^2}$-orbitals.

We can begin to understand the origin of these states by considering the multicentered bonding state formed by Sr's 5s-orbitals within the same atomic layer.
As shown in Fig.~\ref{fig:Sr3CrN3}(d), due to the in-plane 3-fold rotation symmetry, the s-orbitals of Sr atoms that lie in the same plane can form two identical interstitial localized states (labled site-A and site-B).
%, both with a distance of 2.5~\AA away from the cornering atoms.
%Comparing to the predicted critical cage size of 2.73~\AA, this cage provides a very good condition for the electride-like electronic state to form.
Depending on the layer in which these multicentered bonding maxima exist, we label them collectively (the phase difference between them will be addressed later) as s$_\mathrm{bonding}^\mathrm{U/L}$ where U and L are for the upper and lower layer.
However, due to the \emph{interlayer} stacking, only one interstitial site (here, site-A) in each atomic layer is preserved in the bulk structure while forming a 1D interstitial channel.
%However, as we already know, in the end only one of these two maxima is preserved (site-A).
%As it turns out, the other interstitial maxima (site-B) forms another bonding and anti-bonding states with Cr's 3d$_{z^2}$-orbitals in the adjacent layer.

\begin{figure}[ht]
     \centering
     \includegraphics[scale=0.9]{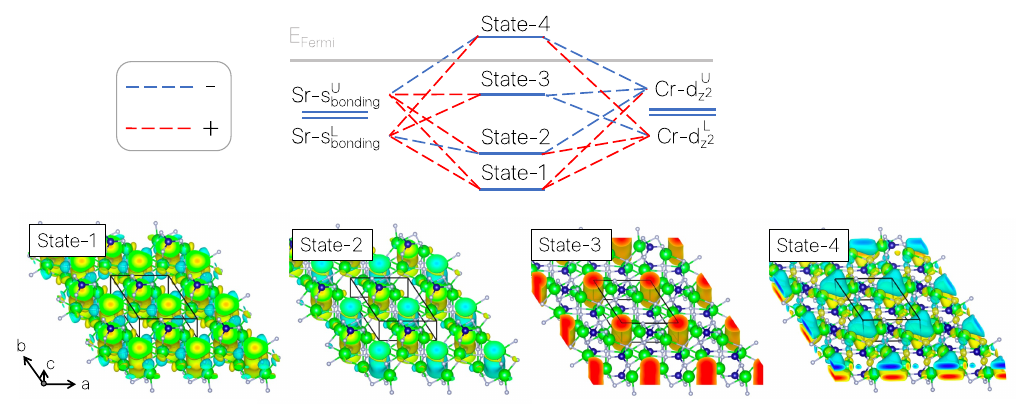}
     \caption{Upper panel: bonding diagram of the Sr$_3$CrN$_3$ system, red dotted lines indicates the atomic orbital has a positive contribution to the molecular orbitals while blue dotted lines indicates a negative contribution. Lower panel: corresponding wavefunction plot at $\Gamma$ point.}
     \label{fig:Sr3CrN3-bonding}
\end{figure}

The other site (site-B) in each atomic layer, however, forms another bonding and anti-bonding pair with Cr's 3d$_{z^2}$-orbitals in adjacent atomic layers.
The bonding state between Cr's 3d$_{z^2}$-orbitals and site-B are found to be state-1 and state-2 [two bands around $-$1.6~eV in Fig.~\ref{fig:Sr3CrN3}(c)] in Fig.~\ref{fig:Sr3CrN3-bonding} where the interstitial localized feature of site-B is destroyed and site-B splits into two sites [see right panel of Fig.~\ref{fig:Sr3CrN3}(b)].
Depending on the phase difference of these resulting bonding orbitals in different layers, the resulting molecular orbitals are labeled as state-1 and state-2 in Fig.~\ref{fig:Sr3CrN3-bonding}.

On the other hand, the anti-bonding states between s$_\mathrm{bonding}^\mathrm{U/L}$ orbitals at site-B and Cr's 3d$_{z^2}$-orbitals in the adjacent layer (state-3 and state-4 in Fig. \ref{fig:Sr3CrN3-bonding}) annihilate the interstitial localized features of s$_\mathrm{bonding}^\mathrm{U/L}$ orbitals at site-B, leaving site-A the only interstitial region left to host electride-like behavior in Sr$_3$CrN$_3$.
Again, depending on the phase difference between layers, these states are further split into bonding [state-3, around $-$0.5~eV in Fig.\ref{fig:Sr3CrN3}(c)], the interstitial band and anti-bonding [state-4, around 4~eV in Fig.\ref{fig:Sr3CrN3}(c)] pairs [see Fig.~\ref{fig:Sr3CrN3-bonding}].

% This bonding/anti-bonding pair causes the energy of state 1 and stat 2 to be much lower than state 3 anad 4 while the energy levels are further splitted by the phase difference between Cr atoms.

% Up until now, we have focused on spread frontier s-orbitals in the main manuscript.
% However, our multicentered bonding theory doesn't prohibits other polarized orbitals from forming such bonds.
% As a matter of fact, A$_5$B$_3$ systems such as Y$_5$Si$_3$ can have occupied frontier d-orbitals forming these interstitials orbitals.
%From the band structure analysis (Fig. ??), one may find that the interstitial orbital should be attributed to the d-orbitals of the Y atoms surrounding the interstitial region.
%However, by putting an onsite Hubbard U of 200 eV on the d-orbitals of Y atoms, we found that these interstitial states lost the features of d-orbitals.
%By putting a U value on Y atoms's 5p-orbitals, we find that these interstitial orbitals energy and shape can be effectively controlled by the U values, indicating that these interstitial states are actually formed by Y atoms' 5p-orbitals.

\clearpage
\section{\label{sec:ELF} Electron localization function}
Proposed by Becke and Edgecombe~\citeS{SBecke1990}, the ELF describes the degree of ``localization'' of electrons by considering the conditional probability of finding an electron (electron $2$) around a point ($\vec r_1$) where one like-spin electron (electron $1$) is located:
\begin{equation}
\begin{aligned}
P^{\sigma\sigma}_\text{cond}\left(\vec r_{1}, \vec  r_{2}\right) \bigg\rvert_{\vec r_{2} \to \vec r_{1}} &= \frac{ P^{\sigma\sigma}\left(\vec r_{1}, \vec r_{2}\right)}{\rho^{\sigma}(\vec r_{1})} \bigg\rvert_{\vec r_{2} \to \vec r_{1}}\\
\label{eq:ELF1}
\end{aligned},
\end{equation}
here, $P^{\sigma\sigma}\left(\vec r_{1}, \vec r_{2}\right)$ is the reduced density matrix which represents the pair probability of simultaneously finding electron 1 with spin $\sigma$ at $\vec r_1$ and electron 2 with spin $\sigma$ at $\vec r_2$, $\rho^{\sigma}(\vec r_1)$ is the one-body charge density of electron 1 with spin $\sigma$.

When $\vec{r}_2$ is close to $\vec{r}_1$, the value of $P_\text{cond}^{\sigma\sigma}\left(\vec r_{1}, \vec r_{2}\right)$ reflects the degree of ``localization" of the electrons at $\vec r_1$: The smaller the conditional probability, the more localized the charge.
As an example, for a point charge at $\vec r_1$ [$\rho^{\sigma}(\vec r_{1})=1$], $P_\text{cond}^{\sigma\sigma}\left(\vec r_{1}, \vec r_{2}\right)$ will be zero, reflecting the perfect localization of a point charge.

Using the fact that when the distance between two electrons is small ($\vec r_{2} \to \vec r_{1}$), only the second and higher order derivatives of the spherically averaged conditional pair density is non-zero~\citeS{SBecke1990}, one can re-express Eq.~\ref{eq:ELF1} as the difference of kinetic energy densities between a Fermionic system and a Bosonic system with the same charge density:
% as the difference between the positively defined kinetic energy density and the so-called von-Weizsacker kinetic energy density [REF]:
\begin{equation}
\begin{aligned}
D_{\sigma}(\vec r) &=\tau_{\sigma}(\vec r)-\frac{1}{4} \frac{\left[\nabla \rho_{\sigma}(\vec r)\right]^{2}}{\rho_{\sigma}(\vec r)}\\
&= 2[\tau_\text{Fermion}(\vec r) - \tau_\text{Boson}(\vec r)].\\
\label{eq:ELF2}
\end{aligned}
\end{equation}
Eq.~\ref{eq:ELF2} gives an alternative interpretation of the ELF: The ELF of a syste reflects the degree of local Pauli exclusion at position $\vec r$.
Covalent bonds usually possess low Pauli exclusion with high charge density, thus, this definition suggests that the ELF can be used to search for occupied covalent bonds.

To assign a more intuitive chemical meaning to $D_{\sigma}(r)$, Becke and Edgecombe~\citeS{SBecke1983} proposed to compare $D$ with its counterpart $D_0$ calculated with a uniform electron gas of the same density at point $\vec r$:
\begin{equation}
\begin{aligned}
\mathrm{ELF}(\vec r)&=\left\{ 1+\left[\frac{D_{\sigma}(\bf r)}{D_{\sigma}^{0}(\bf r)} \right]^2 \right\} ^{-1}\\
D_0({\bf r})&=\frac{3}{5}(6\pi^2)^{2/3}\left[\frac{\rho({ \bf r})}{2}\right]^{5/3}.
\end{aligned}\label{eq:ELF3}
\end{equation}
In this form, the ELF is bounded between 0 and 1 where a number of 0.5 for ELF means the system has the same degree of ``localization'' as the uniform electron gas of the same density and a number of 1 means the electron is completely ``localized'' or that no Pauli exclusion is present at a point in space.

ELF has been previously used to identify the location of the chemical bonds, including multicenter bonds~\citeS{SGrin2014}. 
Naturally, for a system that has an electride-like bond dominating the interstitial area, ELF will likely has a high valued peak at the center of the interstitial region. 
Such feature enables us to use it as a surrogate to identify the existence of the electride-like multicentered bonds.
Extensive studies have also shown that ELF can be used to identify electrides~\citeS{SDale2018, SZhao2016, SWan2018, SZhou2019, SZhang2017prx, SZhang2014, SLee2013, SDong2017nc}.

%\textcolor{red}{[CCX: do we need to compare this to charge density max?]
%Comparing to charge density analysis, ELF analysis is shown to be more robust in searching for the covalent bonds when there are multiple molecular orbitals occupied. 
%This is mainly due to the fact that ELF is a method based on the kinetic energy density of the wavefunctions. 
%A more comprehensive analysis demonstrating the robustness of the ELF in searching for a multicentered bond at the interstitial site can be found in Sec.~\ref{sec:ELF_vs_CHG}.}

Comparing to other wavefunction-based methods like solid state adaptive natural density partitioning~\citeS{SGaleev2013}, maximally localized Wannier functions~\citeS{SMarzari2012}, and crystal orbital bond index~\citeS{SMuller2021} that can also be used to identify multicentered bonds, ELF provides a unique way to identify the location of localized orbitals without relying on unitary transformations of the canonical wavefunctions.
% Okay, it is still a wacefunction basaed method so that this point is reletavely weak as KS orbitals are still being used.

\newpage 
\section{Details of the automatic search protocol}
\label{sec:Search_protocol}
%------------- search protocal ------------------
To search for potential electrides, we have applied our descriptor to the structures in Materials Project (MP) database.
Among the 126,335 structures in the MP database (accessed on 28th November 2021), 51,913 entries were picked for having fewer than 3 types of elements and fewer than 20 atoms in their unit cell. 
Then, DFT calculations were performed for these 51,913 systems (details are described in Sec.~\ref{sec:Methods}) and the charge densities and the ELFs were obtained. 
%And we are ready to perform topological analysis on the ELF to obtain the location of the ELF local maxima and corresponding ELF basin charge.
These files were then analyzed using the descriptor to obtain their likelihood of being identified as electrides. 
%A brief description of the descriptor is provided in the main manuscript (Sec.~\ref{sec:descriptor}) and a more detailed description can be found below.

\begin{figure}[H]
     \centering
     \includegraphics[scale=0.8]{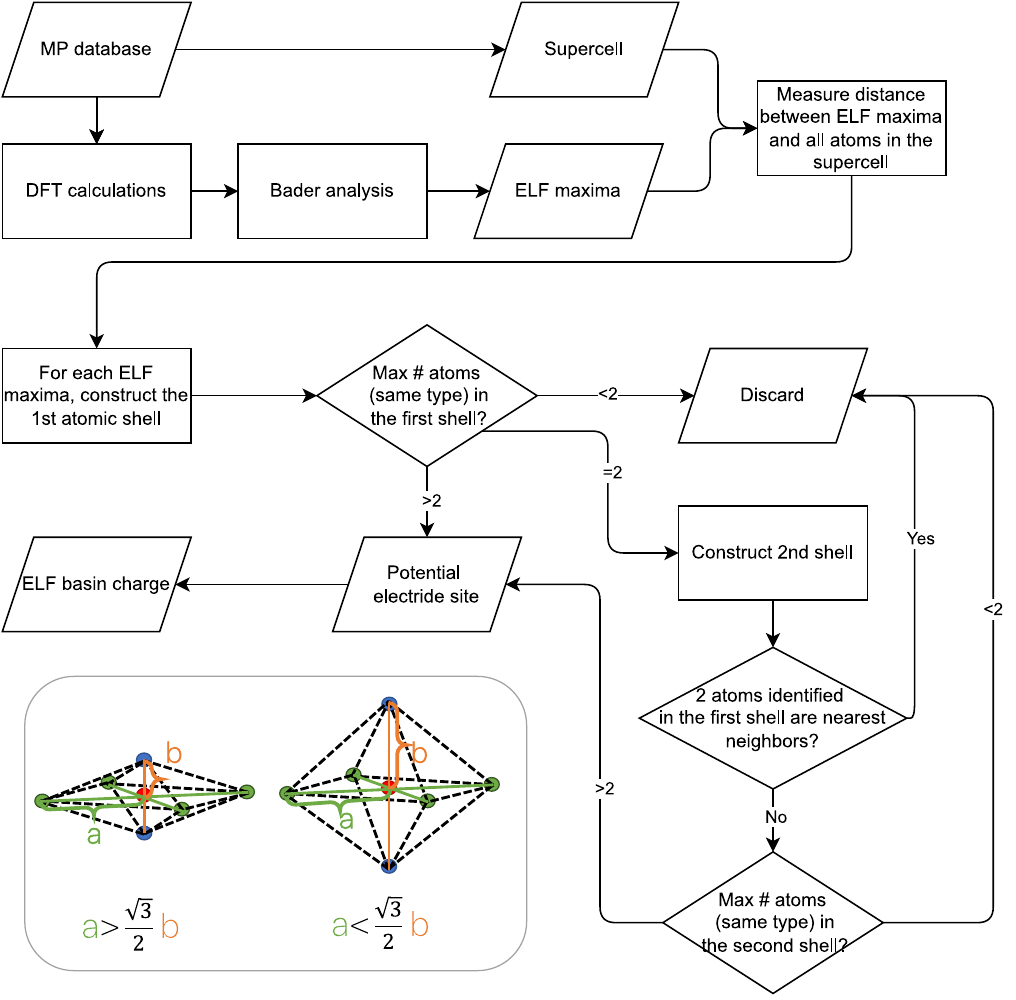}
     \caption{Diagram of the workflow of the screening procedure for potential electrides. Inset: critical condition ($a>\frac{\sqrt{3}}{2}b$) for two atoms found in the first shell (blue atoms) to be nearest neighbors.}
     \label{fig:S-diagram}
\end{figure}

Our electride descriptor starts with performing Bader analysis on the calculated ELF using the Bader code~\citeS{STang2009} and the unit cell is partitioned into basins, each associated with one ELF maxima.
Then, the unit cell is expanded into a $3 \times 3 \times 3$ supercell so that any potential sites located at the edges or faces of the center unit cell are inside the supercell.
For each ELF local maxima in the center unit cell, the distance to all atoms in the supercell are measured and the first shell of atoms is determined using the distance to nearest atom as the inner radius with a shell thickness of 0.2~\AA. 
Atoms that are located inside the first shell are then picked out and, if more than two atoms of the same element are found, we identify this ELF local maximum site as a potential interstitial site.

However, if the maximum number of the atoms of the same type in the first shell is two, then depending on the location of other atoms, they may or may not be nearest neighbors.
% NOTE: these two atoms may well not be the nearest atoms of themselves but what we are concerned of is whether they are nearest neighbor of the atoms in the second shell.
% For non regular cages, this critical condition is treat as a screening parameter.
To determine the critical condition for these two atoms to be nearest neighbors is to measure the distance from the ELF maxima to the nearest second shell of atoms.
As can be seen from the inset of Fig.~\ref{fig:S-diagram}, assuming regular cages, if the distance from the ELF maxima to the atoms in the second shell is larger than $\frac{\sqrt{3}}{2}$ of the distance between them, they are the nearest atom pairs and this site should be discarded.
Hence, we proceed to construct the second shell by choosing the closest atoms to the site that is not in the first shell, and set its distance to the ELF maxima as the the inner radius of the second shell while setting the thickness of the second shell to also be 0.2~\AA.
If the outer radius of the second shell (treated as $a$ in the inset of Fig.~\ref{fig:S-diagram}) is larger than $\sqrt{3}/2 b$, then this site is considered to be a two-center site, otherwise, the same procedure is performed to find the maximum number of atoms in the second shell that are of the same element.
If there are more than two atoms of the same element in the second shell, we also identify this site as a potential multicentered bonding site, otherwise, the site is discarded.
Finally, for the identified sites, the ELF basin charges are obtained by integrating the all-electron charge density over the ELF basins using the Bader code~\citeS{STang2009}. To fit our normalization criteria, any ELF basin charge larger than 2 is reset to 2 (corresponding to interstitial orbital being fully occupied). And the electride figure of merit is calculated using Eq.~\ref{eq:Ze}.
%The electride figure of merit can be calculated using Eq.~\ref{eq:Ze} while resetting the ELF basin charge to 2 for systems with ELF basin charges larger than 2.

For spin polarized systems, same procedure is applied to each spin channel and the ELF maximum is chosen as the one with the highest ELF value.

The ELF is evaluated on the coarse fast Fourier transform grid used to evaluate the wavefunctions. 
Due to the application of the Bader numerical algorithm, one ELF maximum might split into multiple maxima very close to each other. 
This behavior will greatly affect the values of the ELF basin charge.
To avoid this, we apply a merging procedure where the distance between each identified multicentered bonding ELF maxima are measured, and the ones with a distance less than $0.1$ times of the distance to their neighboring atoms are merged.

\clearpage 
\section{Organic electrides}
\label{sec:organic_electrides}
%Organic electrides are the product of attempts aimed at ``freezing" the solvated electrons that appear when ionizing alkaline metals in liquid ammonia.
Organic electrides are usually composed of periodically arranged organometallic complexes (organic molecules encapsulating alkaline metal atoms) with occupied ``Superatom Molecular Orbtials''(SAMOs).
In this section, we'll explore the relation between these SAMOs and the crystal structure of two prototypical organic electrides: Cs(15-Crown-5)$_2$ and Na-Tripip222.

\begin{figure}[ht]
    \centering
    \includegraphics[scale=0.8]{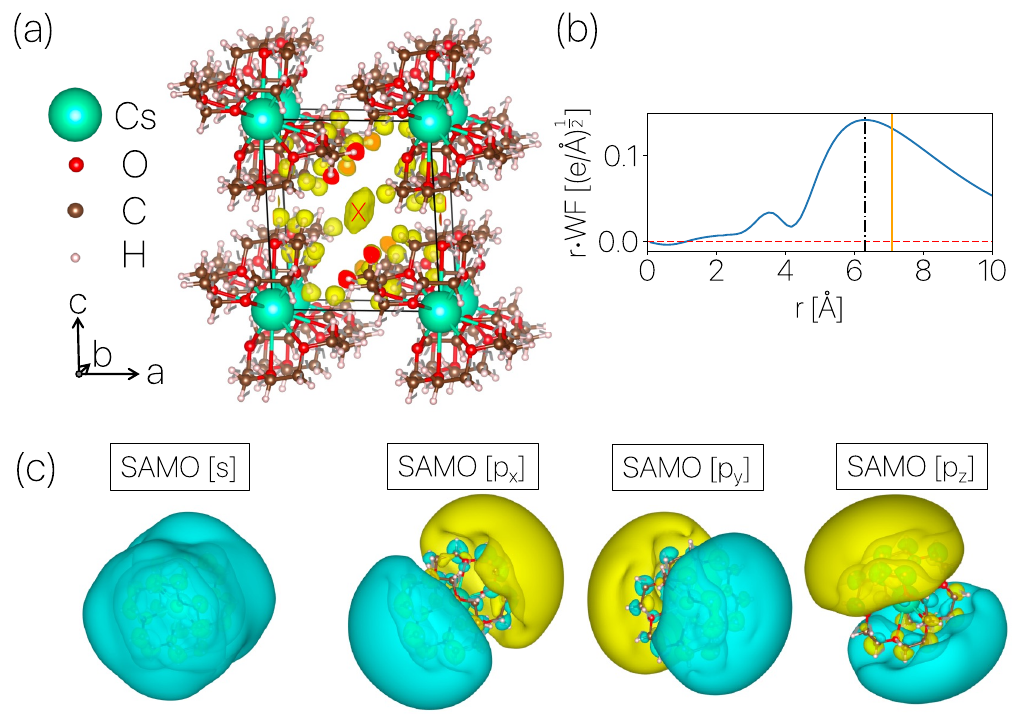}
    \caption{(a) ELF of bulk Cs(15-Crown-5)$_2$, the interstitial site is located at the interstitial region of the unit cell, marked by the red "X". (b) Radial distribution function of the s-shaped SAMO. The maximum point is located at around 6~\AA~(dotted black line) while the distance of the ELF maximum point in the bulk structure to the corner Cs atoms is around 7~\AA~(solid orange line). (c) Wavefunction plot of HOMO (SAMO [s]) and three LUMO (SAMO [p]) of the isolated Cs(15-Crown-5)$_2$ molecule. }
    \label{fig:Cs_crown_ether} 
\end{figure}

We begin by considering the first discovered solid state electride -- cesium crown-ether complex [Cs(15-Crown-5)$_2$].
The crystal Cs(15-Crown-5)$_2$ is composed by four Cs(15-Crown-5)$_2$ complexes at the corners of the unit cell each with two crown ether complexes sandwiching a cesium atom [see FIg.~\ref{fig:Cs_crown_ether}(a)].
The interstitial electrons are localized at the center of the unit cell, occupying a space of approximately 270 ~\AA$^3$~\citeS{SWagner1994,SWard1990,SDawes1991,SDale2014}.
% https://pubs.rsc.org/en/content/articlepdf/2014/cp/c3cp55533j
% https://pubs.acs.org/doi/pdf/10.1021/ja00005a025
% https://journals.iucr.org/c/issues/1990/10/00/cr0150/cr0150.pdf
% https://www.nature.com/articles/368726a0.pdf

The HOMO of the Cs(15-Crown-5)$_2$ molecule is of s-symmetry [see Fig.~\ref{fig:Cs_crown_ether}(c)] and the calculated radial distribution of this s-like SAMO [Fig.~\ref{fig:Cs_crown_ether}(b)] has a peak at around 6~\AA~ with a calculated critical cage size of 8.1~\AA (obtained by fitting the radial wavefunction of the SAMO with hydrogenic 1s-orbital). 
In crystal Cs(15-Crown-5)$_2$, the distance between the interstitial site to the nearest Cs atom is around 6.8~\AA, suggesting that the origin of these interstitial states can potentially be explained as the bonding combination of the SAMOs on the corners of the unit cell.

With our descriptor, we find the ELF maximum value of crystal Cs(15-Crown-5)$_2$ to be 0.99 and the occupation number of the interstitial orbital is 0.6, which gives an electride figure of merit $Z_\mathrm{e}=0.65$. 
This result suggests that the interstitial state is only partially occupied and that electron doping or changing Cs to Ba can potentially be used to increase the likelihood of it being identified as an electride.

\begin{figure}[H]
     \centering
     \includegraphics[scale=0.8]{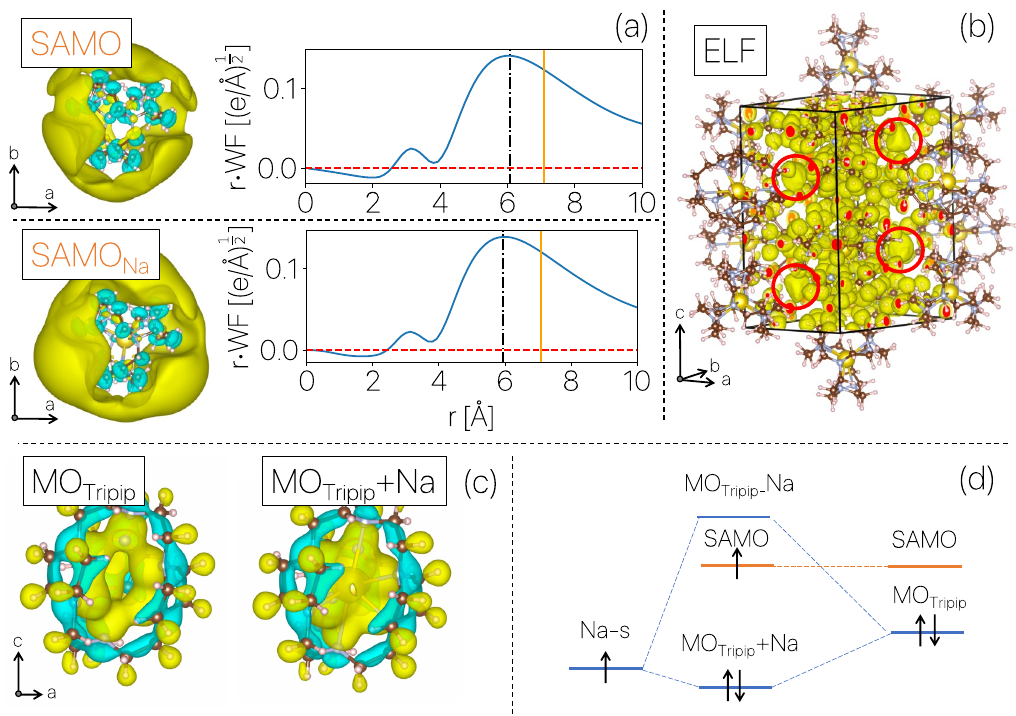}
     \caption{(a) Wavefuntion and radial distribution function plot of the s-shaped SAMOs of the Tripip222 molecule with (lower panel) and without (upper panel) Na atom at the center. Dotted black line indicates the location of the outer most maxima in the RDF and the yellow line indicates the distance of the center of the interstitial state in the bulk phase to the center Na atoms of the surrounding organometallic complexes. (b) ELF plot of the electride phase of Na-Tripip222. (c) Wavefunction plot of the doubly occupied MO$_\text{Tripip}$ state before and after hybridizing with Na's s-orbital. (d) Bonding diagram of the Na-Tripip222 organometallic molecule.}
     \label{fig:Na-tripip}
\end{figure}

As another example, organic electride -- crystal Na-Tripip222 is composed by four Na-Tripip222 molecular building blocks in the unit cell. 
As can be seen in Fig.~\ref{fig:Na-tripip}(a), without the Na atom, the LUMO of the building block -- Tripip222 organic molecule has an unoccupied s-shaped SAMO.
Surprisingly, the insertion of Na atom has limited effect on the SAMO of the Tripip222 molecule [Fig.~\ref{fig:Na-tripip}(a), lower panel] and merely acts as an electron donor.
This is because Na's 3s-orbital hybridizes with another low energy molecular orbital of Tripip222 that was already fully occupied [see Fig.~\ref{fig:Na-tripip}(c)], which, in turn, leads to the occupation of the s-like SAMO [Fig.~\ref{fig:Na-tripip}(d)].

In crystal from, Na-Tripip222 has a high valued ELF maximum at the center of the interstitial regions between the organometallic complexes, as showing in Fig.~\ref{fig:Na-tripip}(b).
The distance between these maxima and the closest Na atoms is $\sim$ 7~\AA, comparable to the calculated critical cage size $b_c$ of s-like SAMO of 7.66~\AA (obtained by fitting the radial wavefunction of the SAMO with hydrogenic 1s-orbital), suggesting that the interstitial electronic state in Na-Tripip222 could be potentially attributed to the linear combination of the SAMOs of the organometallic molecules.
%By projecting wavefunction onto the s-like SAMOs, we find 4 bands are mainly composed of these SAMOs (see Fig. \ref{}) and their wavefunction plot at $\Gamma$ indeed demonstrate that they are the electride-like states, confirming the applicability of our multicentered bonding theory.

\newpage
\section{F-center defects}
\label{subsec:other_electrides;F-center}

Although usually discussed in crystal systems, the ``electride-like'' interstitial electronic states have also been found in non-periodic structures.
The F-center (or color-center) defects for example, also exhibit this behavior.
The F-center defect is generated when an anion is removed from the crystal.
After the anion is removed, the remaining electrons are localized at the vacancy site and form an interstitial localized state.
Previous studies have pointed out that these states are very similar to the electride-like states~\citeS{SPopov2018,SKulichenko2020}.
%https://pubs.acs.org/doi/full/10.1021/acs.jpcc.8b03118
%https://www.sciencedirect.com/science/article/pii/S0301010419314685
%A whole bunch of Hosono's work! and Alexander I. Boldyrev's work

\begin{figure}[ht]
    \centering
    \includegraphics[scale=0.8]{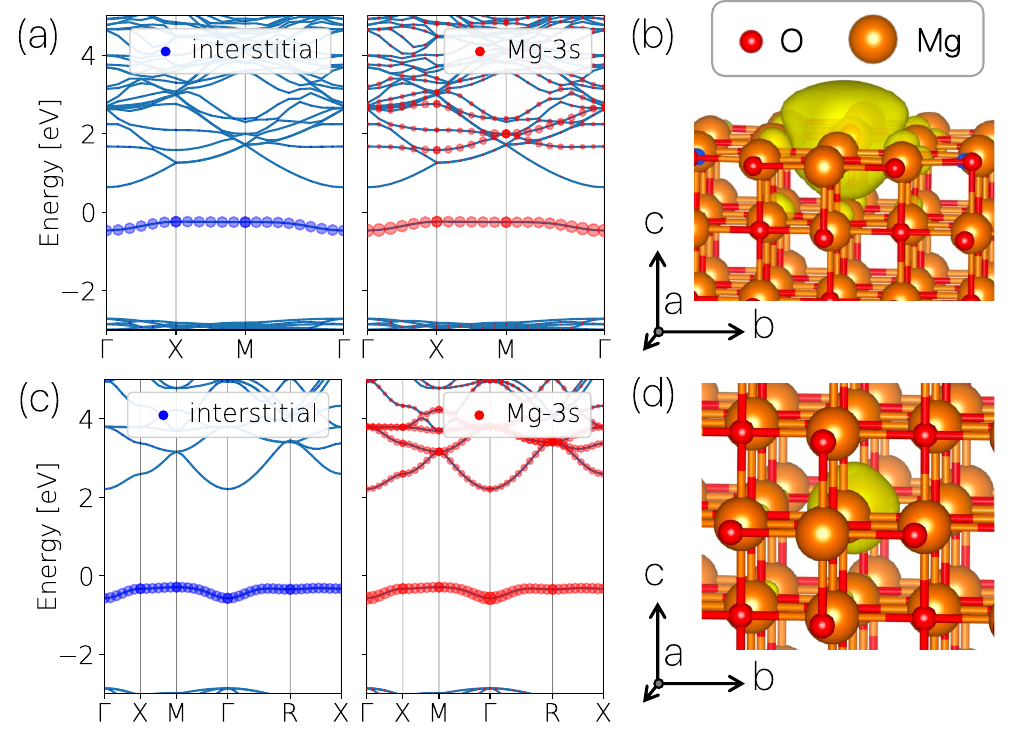}
    \caption{(a) Projected band structure of [001] surface of MgO with an oxygen defect. The size of the blue and red circles indicates the contribution of the interstitial orbital (by projecting onto a hydrogenic 1s orbital at the defect site) and surrounding Mg's 3s-orbitals, respectively. (b) Charge density related to the defect band of oxygen defect on the [001] surface of MgO. (c, d) Same as (a, b) but the oxygen defect is located inside the bulk MgO.}
    \label{fig:MgO} 
\end{figure}

For example, a F-center defect in MgO is generated when an oxygen atom is removed.
If this oxygen vacancy (V$_\mathrm{O}$) is located at the surface, one finds a fully occupied defect state that's isolated from the rest of the band structure [see Fig.~\ref{fig:MgO}(a)].
The charge density corresponding to the defect band [Fig.~\ref{fig:MgO}(b)] suggests that it is an interstitial localized state.
Projected band structure indicates that this defect band can be described as a bonding combination of the five surrounding Mg atoms' 3s-orbitals [see Fig.~\ref{fig:MgO}(a), right panel].
%by the surrounding five Mg atoms' s-orbitals with an average projection coefficient of \textcolor{red}{XXX} while the bottom Mg's 3s-orbital has a slightly higher projection coefficient of \textcolor{red}{XXX}.
Similarly, if the V$_\mathrm{O}$ is located inside the bulk, then the interstitial state should be attributed to the six surrounding Mg's 3s-orbitals [as indicated by the projected band structure shown in Fig.~\ref{fig:MgO}(c,d)]
%by six surrounding Na atoms' s-orbitals with a projection coefficient of \textcolor{red}{XXX} on each s-orbitals. 
This findings coincide with the previous results~\citeS{SPopov2018} where analysis was performed using the NBO and SSAdNDP, confirming the multicentered bonding feature of this state.

For both surface and bulk oxygen defects, the measured cage size ($b$) is around 2.12~\AA, comparable to the calculated critical cage size of Mg ($b_\mathrm{c}^\mathrm{[3D]}=2.17$~\AA), suggesting that the interstitial orbital responsible for the electride-like behavior could be described as a multicentered bonding combination of surrounding Mg's 3s-orbitals.
%to the radius of Na's frontier s-orbital (\textcolor{red}{XXX}Å) this provides a perfect condition for an electride-like state to form.
%When the O$_s$ defect is on a clean MgO [001] surfrace, the defect site looses one of its six mirror symmetries, suggesting that the center of the electride-like state should deviate from the oxygen vacancy site, which is also observed in the projected charge density plot of the defect band (Fig.~\ref{}). 

%F-center defects can be seen as the dilute limit of an electride where the electride-like states are faraway from each other.
%By increasing the concentration of the F-center defects, one can reach the electride-like limit.
%i.e. Electrides can be understood as systems with periodically arranged F-center defects.
%In a hypothetical situation, take MgO as an example, this corresponds to removing all O atoms in the system wile keeping the positions of Mg atoms unchanged.
%Fig.~\ref{fig:results17} demonstrates this hypothetical situation, from the ELF plot on Fig.~\ref{fig:results17}, we can clearly see that there are occupied interstitial electride-like orbitals in this system.
%Applying our descriptor, we get an max interstitial ELF amplitude of  \textcolor{red}{XXX} with an occupation of \textcolor{red}{XXX}, putting it well within the electride-like region.

\newpage 

\bibliographystyleS{apsrev4-2}
\bibliographyS{bibliography/SI}

\end{document}